\documentclass[hyper]{JHEP3}

\input{epsf}
\usepackage{epsfig}
\usepackage{amssymb}
\usepackage{amsfonts}
\usepackage{amsbsy}
\usepackage[all]{xy}
\usepackage{amsmath}

\usepackage{amssymb,amscd}
\usepackage{mathrsfs}
\usepackage{amsmath,amsthm}
\def\be{ \begin{equation} }
\def\ee{ \end{equation}}

\def\Aut{{\rm Aut}}
\def\ch{{\rm ch}}

\def\cot{{\rm cot}}

\def\dim{{\rm dim}}
\def\End{{\rm End}}
\def\exp{{\rm exp}}

\def\Hom{{\rm Hom}}
\def\I{{\rm i}}

\def\ker{{\rm ker}}
\def\log{{\rm log}}
\def\Mat{{\rm Mat}}

\def\Tr{{\rm Tr}}

\def\vol{{\rm vol\,}}

\def\half{\frac{1}{2}}
\def\p{\partial}

\def\one{{\hbox{ 1\kern-.8mm l}}}

\def\CA{{\cal A}}
\def\CB{{\cal B}}
\def\CC {{\cal C}}
\def\CD {{\cal D}}
\def\CE {{\cal E}}
\def\CF {{\cal F}}

\def\CH {{\cal H}}

\def\CK {{\cal K}}

\def\CN {{\cal N}}
\def\CO {{\cal O}}

\def\CR {{\cal R}}

\def\CW {{\cal W}}

\def\CO {{\cal O}}

\def\CE {{\cal E}}

\def\CH {{\cal H}}

\def\CB {{\cal B}}

\def\CS {{\cal S}}

\def\IB{\mathbb{B}}
\def\IC{\mathbb{C}}

\def\IP{\mathbb{P}}

\def\IR{{\mathbb{R}}}

\def\IT{{\mathbb{T}}}

\def\IZ{{\mathbb{Z}}}

\def\fs{\mathfrak{s}}

\def\fs{\mathfrak{s}}

\def\fu{\mathfrak{u}}

\def\fA{\mathfrak{A}}
\def\fB{\mathfrak{B}}

\def\fM{\mathfrak{M}}

\def\rmk#1{\bigskip\noindent{\bf Remarks} }

\def\BR{{\rm BR}}
\def\EV{{\rm EV}}

\title{Quantum Mechanics With Noncommutative Amplitudes}

\author{Gregory W.~Moore  \\
  NHETC and
Department of Physics and Astronomy, Rutgers University \\
126 Frelinghuysen Rd., Piscataway NJ 08855, USA\\
{\tt gmoore@physics.rutgers.edu} }

\abstract{We formulate a Born rule for families of
quantum systems parametrized by a noncommutative space of
control parameters. The resulting formalism may be viewed
as a generalization of quantum mechanics where overlaps take values in a
noncommutative algebra, rather than in the complex
numbers.
\today }

\begin{document}

\section{Introduction And Conclusion}\label{sec:Introduction}

Control parameters are ubiquitous in physics. Usually the space $X$
of control parameters for a family of quantum mechanical systems is a
topological space (possibly disconnected). In this paper we
discuss what it would mean to replace $X$ by a noncommutative
space. We will be lead to a formalism
which can, if one is so disposed,  be interpreted as a slight generalization of quantum
mechanics.

In colloquial speech physicists generally refer to a
quantum state as a nonzero vector $\psi$ in a complex Hilbert space.
At this colloquial level,  a continuous family of quantum
states is a continuous family $x \mapsto \psi_x$ of nonzero vectors,
 where $x \in X$ runs over the control parameters.
  Such a continuous family is simply a continuous section $\Psi$ of a bundle of
Hilbert spaces over $X$.   Now we want to replace $X$
by a ``noncommutative space,'' whose algebra of functions is a, possibly
noncommutative, $C^*$-algebra $\fA$, and to replace continuous families of vectors
$\psi_x$ by a single vector $\Psi$ in a Hilbert $C^*$-module $\CE$ over $\fA$.
(See section \ref{sec:HilbMod}  for a definition of a Hilbert $C^*$-module.)
The ``inner-product'' on two such  vectors $\Psi_1$ and $\Psi_2$,
denoted $\langle \Psi_1, \Psi_2\rangle$,  is valued in $\fA$,
and therefore we refer to this generalization of quantum mechanics
as ``quantum mechanics with noncommutative amplitudes,''
or QMNA for short.

In QMNA matrix elements of self-adjoint operators  $T$ such
as $\langle \Psi_1, T \Psi_2\rangle $ are elements of a
possibly noncommutative algebra $\fA$ rather than
complex numbers. Therefore, the statement of the Born
rule for computing probability amplitudes is not immediately
obvious. We will elaborate on this point in section \ref{sec:QMNA-BR}.  The main
goal of this paper is to suggest a sensible statement of the Born rule in this context.
This is the content of equation \eqref{eq:BR-NC}.
We will also attempt to address two crucial questions:

\begin{enumerate}

\item  First, one may well wonder if QMNA really is a generalization
of quantum mechanics, or just standard QM dressed up in a
fancy Halloween costume.

\item Second, while the
structure we investigate seems (at least, to the author) to be
 mathematically well-motivated it is not
clear whether or not it has any compelling physical applications.

\end{enumerate}

The answer to the first question is ``Yes and no.'' It is
best discussed, (see section \ref{sec:ReallyGeneralize})  after the detailed proposal has been made
and some examples have been examined.  Regarding the second question our answer is - at least for the moment - ``Not yet."
Nevertheless, there are several potential   applications, among them:

\begin{enumerate}

\item Whenever  the space of control parameters is a symplectic manifold $X$ we
can try to quantize it. We discuss this in detail in section \ref{sec:Semiclassical}
in the special case when $X$ is an algebraic K\"ahler manifold. The procedure has some
overlap with the Born-Oppenheimer approximation, but should be
distinguished from it.

\item The famous noncommutative torus has many applications in physics. We recall some of them in
section \ref{subsec:PhysicalInterpret} since each instance provides a potential route for applications
of this paper.

\item Moduli spaces of vacua in supersymmetric field theories offer a wide variety of families of
quantum systems. In much recent literature the quantization of these spaces has been discussed
\cite{Beem:2016cbd,Bullimore:2015lsa,Bullimore:2016nji,Cecotti:2013mba}.
These, or related, considerations might provide a framework for application of the ideas of this paper,
although it must be admitted that the authors of these papers will find the kinds of questions
addressed here somewhat alien.
Similar considerations might hold for moduli spaces of string compactifications.

\item If a space of control parameters has an ergodic action by a group then it is natural to consider
the space of quantum systems over the ``quotient space.'' Such quotient spaces are canonical sources
of noncommutative manifolds \cite{Connes2,Varilly:1997qg}. As an example, $T$-duality acts ergodically on the boundary of Narain
moduli space. (This is closely related to facts exploited in \cite{Moore:1993zc}.)

\item Finally, and most grandiosely, the ultimate control parameters are the parameters of the
landscape.   Perhaps there is a notion of a ``noncommutative landscape.'' For example in the
context of string theory flux compactification \cite{Denef:2007pq,Douglas:2006es} one could
imagine replacing the discrete points parametrizing fluxes  in flux compactification by fuzzy points.
(This would entail a change in the mathematical description of RR fields using differential
cohomology \cite{Belov:2006xj,Freed:2006ya,Freed:2006yc}.) If sensible, then our considerations
might have a bearing on the notorious ``measure problem'' of cosmology.
\footnote{And once one goes down this path it is necessary to ask if different ``parts'' of
the ``multiverse'' are even based on the same $C^*$ algebras.}

\end{enumerate}

As a sanity check on our proposal we examine several examples:
In section \ref{sec:FiniteDimensional} we consider the general case of
finite-dimensional $C^*$-algebras. Curiously, some well-known formulae
from quantum information theory arise naturally. Thus, for what it is worth,
one could rephrase at least some statements in quantum information theory
in terms of noncommutative geometry. One famous way to quantize a space
is in the case when $X$ is a K\"ahler manifold equipped with a very ample
holomorphic Hermitian line bundle. We explore what can be said in this
context in section \ref{sec:Semiclassical}. (When $X$ is compact this
is really a special case of the
finite-dimensional setup of section \ref{sec:FiniteDimensional}, but the extra
geometrical structures allow us to discuss new aspects such as a semiclassical
limit.) In section \ref{sec:NC-Torus} we consider a very famous noncommutative
manifold - the noncommutative torus.

Most of this note addresses kinematic issues.
If the proposal of this paper makes any sense, then it is important to consider dynamical issues and the role
of symmetry. We begin this investigation with some brief remarks in
section \ref{sec:TimeDevelopment}. We also discuss more general symmetries than
time translation, and the extent to which the usual Wigner theorem applies.
We will argue that our setup provides a framework in which one can realize
Weinberg's idea \cite{Weinberg:2014ewa} that symmetries can be implemented in a way more general than
envisaged in the Wigner theorem.

Finally, in Appendix \ref{app:Theorems} we review some important technical constructions
and work them out for the case of finite-dimensional $C^*$-algebras.

\section*{Acknowledgements}

This project has its origin in an unpublished section of the paper \cite{Freed:2012uu}.
I would like to thank C. Akemann, T. Banks, J. Bellissard, S. Donaldson,
D. Freed, D. Friedan, J. Fr\"ohlich, D. Harlow, M. Hastings, A. Kapustin, A. Kitaev T. Mainiero,
J. Maldacena, W. Peelaers, N. Read, G. Segal, and E. Witten for helpful discussions and correspondence. I am very grateful to J. Bellissard
for sharing his unpublished notes on completely positive maps \cite{BellissardCP}. I thank T. Banks, D. Harlow, and especially T. Mainiero for useful comments on the draft.  Some of this paper was written while visiting the Institute for Advanced Study in Princeton and I am very
grateful to the IAS for hospitality and support from   the IBM Einstein Fellowship of the Institute for Advanced Study. Some of this work was done at the Aspen Center for Physics  (under NSF Grant No. PHY-1066293). I also acknowledge support by the DOE under grant
DOE-SC0010008 to Rutgers.

\section{Quantum Systems}\label{sec:QuantumSystems}

In this paper we will have nothing to say about the deep
problems of measurement theory: We accept hook, line, and sinker
the framework laid out, for example in \cite{Dirac,Holevo}.
(See \cite{BanksBook,Blanchard:2016muv,Frohlich1,Mermin:1993zz, WeinbergTextBook} for a small sampling of useful discussions
in this huge literature.)
By a ``quantum system'' we simply mean a mathematical definition of
a notion of ``physical states,'' ``physical observables,''
and a pairing ${\rm BR}$ of such states and observables to give a
probability distribution on the real numbers. Given a state $s$ and
observable $O$ we get a probability distribution ${\rm BR}(s,O)$.
The probability ${\rm BR}(s,O)(m)$ associated to a measurable subset $m\subset \IR$
 is meant to reflect the intrinsic probability that a measurement of the observable
 $O$ in the state $s$ takes its value in the set $m$.
We will refer to such a pairing as a \emph{Born rule}.

In standard quantum mechanics these notions are made precise
by the very well-known Dirac-von Neumann axioms:

\begin{enumerate}

\item  Physical observables are identified
with self-adjoint operators $T$ on a complex Hilbert space $\CH$.
 Because we wish to avoid technicalities of functional analysis
as much as possible we will restrict attention to bounded operators
on a separable Hilbert space.   We denote the space of bounded operators by $\CB(\CH)$. We will
denote the set of physical observables by $\CO$ so
in ordinary quantum mechanics, $\CO := \CB(\CH)_{s.a.}$.

\item
Physical states are identified with ``density matrices,'' that is,
with positive traceclass operators $\rho$ on $\CH$ with $\Tr_{\CH}(\rho)=1$.
The space of physical states is denoted by $\CS$.

\item  Time evolution is defined by a continuous family of unitary automorphisms of the
space of physical states, or, equivalently, by a continuous family of
unitary automorphisms of the space of observables.

\item Symmetries are identified with homomorphisms of a group into the group of automorphisms
of the Born rule. The latter are determined by a map $\alpha: \CS \rightarrow \CS$
 that is affine linear on convex combinations: $\alpha( t_1 \rho_1 + t_2 \rho_2) =
t_1\alpha(\rho_1) + t_2 \alpha(\rho_2)$ for any two states $\rho_1, \rho_2 \in \CS$ and
$0\leq t_1, t_2$ such that $t_1 + t_2 =1 $. Using results from \cite{Kadison} one
can show that $\alpha(\rho) = U \rho U^*$ where $U$ is a unitary or antiunitary operator on $\CH$.

\end{enumerate}

The physical content is encoded in the Born rule. We will regard the Born
rule as a map
\be
\BR:  \CS\times \CO \to \fM(\IR)
\ee
where $\fM(\IR)$ denotes the set of probability measures on $\IR$.
To define the map $\BR$ we need the spectral theorem. The spectral
theorem states that there is
a one-one correspondence between bounded self-adjoint operators $T\in \CB(\CH)$
and projection-valued measures. If $P_T(m)$ denotes the projector associated
to self-adjoint operator $T$ and measurable set $m\subset \IR$ then
 the probability distribution $\BR(\rho,T)$ is defined by:
\be\label{eq:StandardBR}
\BR(\rho,T)(m):= \Tr_{\CH} \rho P_T(m)\in \IR_+ .
\ee

As a special case of \eqref{eq:StandardBR} suppose that $T$ has discrete spectrum so that
\be
T = \sum_{\lambda\in \sigma(T)}  \lambda  P_{\lambda }
\ee
where $\sigma(T)$ is the spectrum of $T$ and
$P_{\lambda }$ are projectors onto the   eigenspaces for
the distinct eigenvalues of $T$.  Then
\be
P_T(m) = \sum_{\lambda \in m}  P_{\lambda}.
\ee
Now, if $\rho$ is a pure state given by a rank one projector, usually
written as
$\rho = \vert \psi \rangle \langle \psi \vert$ in bra-ket notation,
then to find the probability
of measuring exactly $\lambda_0$ as the value of $T$ in the state $\rho$
we take the set $m$ to be an interval containing $\lambda_0$ and no other
eigenvalues. Then \eqref{eq:StandardBR} reduces to the standard formula
from elementary textbooks:
\be
\BR(\rho,T)(m)= \langle \psi , P_{\lambda_0}  \psi \rangle.
\ee
  If we choose an orthonormal basis $\psi_{\lambda_0}^a$ for
the image of $P_{\lambda_0}$ this can be further written as:
\be\label{eq:QM-text}
\BR(\rho,T)(m)= \sum_a  \langle \psi, \psi^a_{\lambda_0} \rangle \langle \psi, \psi^a_{\lambda_0} \rangle^*.
\ee

It is often simpler to work with expectation values
 $\EV: \CO \times \CS \rightarrow \IR$. These are
 related to the Born rule by $ \BR(T,\rho)(m):= \EV(P_{T}(m), \rho) $.

\subsection{Continuous Families Of Quantum Systems}\label{sec:ContFamQM}

Except for some issues of choosing suitable operator topologies,
which are discussed at length in Appendix D of \cite{Freed:2012uu},
the generalization of the previous discussion to
 continuous families of quantum systems is completely
straightforward: Let $X$ be a topological space, as in the Introduction.
We introduce a Hilbert bundle $\pi: \CH \to X$ and denote the
Hilbert space fiber above a point $x\in X$ by $\CH_x$. Then there is a
bundle of algebras of bounded operators $\CB(\CH)$.  The set $\CO$ of
``observables''
is now the set of continuous sections of $\CB(\CH)$, namely, a continuous family
$x \mapsto T_x \in \CB(\CH_x)$. Similarly the set $\CS$ of ``states'' is the set of continuous
sections of positive traceclass operators with $\Tr_{\CH_x}(\rho_x) = 1$.
We now have a continuous family of probability distributions $\wp_x$ parametrized by $x\in X$.
So, for every measurable subset of $\IR$ we have:
\be
\wp_x: m \mapsto  \Tr_{\CH_x}(\rho_x P_{T_x}(m))
\ee
Therefore we should regard the ``Born rule'' as a map
\be\label{eq:Family-BR-1}
\CS \times \CO \times X \to \fM(\IR)
\ee
so that ${\rm BR}(\rho,T,x)=\wp_x$.

\section{Hilbert $C^*$-Modules}\label{sec:HilbMod}

We now seek to generalize the previous section by following the standard philosophy of
noncommutative geometry \cite{Connes2}: We rephrase statements about $X$ in terms of
its $C^*$-algebra of functions $C(X)$, as in the Gelfand theorem. Next we contemplate
the analogous statements obtained by replacing   $C(X)$
by a general  (possibly noncommutative)  $C^*$-algebra $\fA$.
In the discussion that follows we will often motivate statements and constructions
by returning to the case $\fA = C(X)$. We call this the ``commutative case'' for
brevity.

Next, following standard noncommutative geometry, we replace $\CH$ by
its space of sections $\Gamma(\CH)$ and then replace this
by a right (finitely generated and projective)
$\fA$-module $\CE$.
\footnote{A \emph{right $\fA$-module} is a vector space $\CE$ with a right-action
$\CE\times \fA \to \CE$ denoted $(\Psi, a) \rightarrow \Psi\cdot a$ that is
compatible with the linear structures on $\CE$ and $\fA$. A projective $\fA$ module is
a direct summand of $\fA^{\oplus n}$ for some $n$. Put differently, it is a module
of the form $P \cdot \fA^{\oplus n}$ where $P$ is an $n\times n$ projection
operator in $M_n(\fA)$. }
The extra structure of a Hilbert space is
generalized to the statement that $\CE$ is a \emph{Hilbert $C^*$-module
over $\fA$}, a notion we will define presently.

To motivate the definition of a Hilbert $C^*$-module note that
in the commutative case a continuous section $\Psi: x \mapsto \psi_x \in \CH_x$ of
$\CH$ gives a continuous function
\be
x \mapsto \langle \psi_x, \psi_x \rangle
\ee
and hence there is a nondegenerate sesquilinear product on $\Gamma(\CH)$ valued in
the $C^*$-algebra $C(X)$:
\be
\langle \Psi,\Psi\rangle \in C(X).
\ee
This motivates the definition \cite{Paschke,Rieffel-0}. (For
a sampling of literature see \cite{Connes2,Lance,Landsman,Manuilov,RaeburnWilliams,WeggeOlsen}.
\footnote{ In thinking about this project we have found the expository
notes by Landsman \cite{Landsman} to be especially clear, relevant, and helpful.
See also \cite{LandsmanBook}.})

\bigskip
\noindent
\textbf{Definition} A \emph{Hilbert $C^*$-module
over a $C^*$-algebra $\fA$} is a right $\fA$-module,
denoted $\CE$,  with a nondegenerate positive
 $\fA$-valued inner product. That is:

a.)  For all vectors $\Psi_1, \Psi_2 \in \CE$ we have
\be\label{eq:HM-1}
\langle \Psi_1,\Psi_2 \rangle\in \fA
\ee
It is $\IC$-antilinear in the first and $\IC$-linear in the second argument.

b.) The inner-product is compatible with the $*$-structure
\be\label{eq:HM-2}
\langle \Psi_1,\Psi_2 \rangle^* = \langle \Psi_2,\Psi_1 \rangle
\ee
as well as with the $\fA$-module structure:
\be\label{eq:HM-3}
\langle \Psi_1,\Psi_2 a\rangle = \langle \Psi_1,\Psi_2 \rangle a
\ee

c.) For all $\Psi \in \CE$,
\be\label{eq:HM-4}
\langle \Psi,\Psi \rangle \in \fA_+
\ee
is a positive element in $\fA$. Moreover it vanishes iff $\Psi =0$.

d.) $\CE$ is a complete normed vector space (that is, a Banach space) with respect to the norm
\be\label{eq:HM-5}
\parallel \Psi \parallel := \sqrt{ \parallel \langle \Psi,\Psi \rangle \parallel_{\fA} }
\ee
(where $\parallel \cdot \parallel_{\fA}$ is the norm on $\fA$).

\bigskip
\noindent
\textbf{Remarks}:

\begin{enumerate}

\item
Note that it easily follows from \eqref{eq:HM-2} and \eqref{eq:HM-3} that
 $\langle \Psi_1 a ,\Psi_2 \rangle = a^* \langle \Psi_1,\Psi_2 \rangle $.

\item We stress that while $\CE$ is a Banach space it need not be a
Hilbert space. The norm on $\CE$ need not satisfy the parallelogram
law and there are indeed  examples of
Hilbert modules that are not Hilbert spaces in any natural way. For example,
any $C^*$ algebra $\fA$ is a $C^*$-module over itself with $\langle a_1, a_2\rangle:= a_1^* a_2$.
 Of course, there are also
examples of Hilbert modules that naturally admit a Hilbert space structure, and we will
see many examples below

\item One aspect of Hilbert space theory that does not generalize is that
if $T: \CE \to \CE$ is a bounded linear operator it does not
immediately follow that there is an adjoint. Let us  define \emph{adjointable operators} to be
those $\IC$-linear transformations $T:\CE \to \CE$
 such that there exists $T^*: \CE \to \CE$ such that
for all $\Psi_1, \Psi_2$
\be
\langle T^* \Psi_1, \Psi_2\rangle = \langle \Psi_1 , T \Psi_2\rangle
\ee
A simple argument shows that such adjointable operators are
``$\fA$-module maps,'' meaning:
\be\label{eq:A-module-map}
T(\Psi a) = T(\Psi) a
\ee
for all $\Psi \in \CE $ and $a\in \fA$. The necessary condition
\eqref{eq:A-module-map} cuts down the set of linear maps on $\CE$ substantially,
and in particular, not all linear operators in $\End(\CE)$ are adjointable. One can show that
the set of all adjointable operators, denoted $C^*(\CE, \fA)$,
is itself a $C^*$-algebra (using the operator norm) acting on $\CE$ from the left.
A particularly important set of adjointable operators are the operators $T_{\Psi_1, \Psi_2}$
defined by
\be
T_{\Psi_1, \Psi_2}(\Psi_3):= \Psi_1 \langle \Psi_2, \Psi_3 \rangle
\ee
The operators $T_{\Psi_1, \Psi_2}$ are analogs of finite-rank operators and
generate a norm-closed subalgebra of $C^*(\CE,\fA)$ denoted
$C^*_0(\CE,\fA)$ that, in some sense, play the role of compact operators in Hilbert space
theory.

\item A   second aspect of Hilbert space theory that does not generalize is
orthogonal decomposition with respect to closed subspaces. If
$\CW \subset \CE$ is a closed subspace one can define the orthogonal complement:
\be
\CW^\perp: = \{ \Psi \vert \langle \Psi, \Psi' \rangle =0 \qquad \forall \Psi' \in \CW \}.
\ee
However, it is not necessarily true that
$\CE = \CW \oplus \CW^\perp$.  As a simple example, we again take $\CE = \fA$ to be a Hilbert
module over itself with $\fA$-valued inner product
$\langle a_1, a_2 \rangle = a_1^* a_2$. Now let
$\fA = C(X) = \CE$   for some topological space $X$ and
let $Y\subset X$ be a closed subspace whose complement
is dense. Now take $\CW$ to be the submodule of $\CE$
defined by functions that vanish on $Y$. This is a closed
submodule: The limit of continuous functions that vanish on $Y$
will vanish on $Y$.
On the other hand, any function in $\CW^\perp$ would have
to vanish on an dense subset of $X$ and would therefore have
to vanish. Therefore $\CW^\perp = \{ 0 \}$, so
$(\CW^\perp)^\perp = \CE$.

\item It would be interesting to see if there are any physical implications
of the technicality discussed in the previous remark.
For example, it follows that the lattice of subspaces of a Hilbert module
is \underline{not} an orthocomplemented lattice so that the ``quantum logic''
of \cite{BirckhoffvonNeumann,Mackey} does not apply.
However, it follows from Lemma 15.3.4 of \cite{WeggeOlsen} that the set of
``complementable subspaces'' of a Hilbert module is an orthocomplemented lattice.
A subspace $\CW$  is ``complementable'' if $\CE = \CW \oplus \CW^\perp$.
Equivalently, if it is the range of an adjointable projection operator
$P \in C^*(\CE,\fA)$. It would be interesting to find a generalization
of Gleasons' theorem  \cite{Gleason} in this context.
\footnote{Already the first step of Gleason's argument does not generalize
to the Hilbert module case. He reduces the probability function to a function
on one-dimensional subspaces using additivity under orthogonal sum. But
even for $M_n(\IC)$, considered as a Hilbert module over itself, there is
no vector space basis of orthogonal elements.}

\end{enumerate}

\section{Observables And States}\label{sec:ObsStates}

We will now define observables and states for QMNA. For simplicity we henceforth assume the
$C^*$ algebra $\fA$ to be unital.

The set of (bounded) observables $\CO$ is straightfoward to generalize:
We simply take the set of observables to be the set $\CO(\CE,\fA)$ of
 self-adjoint elements in $\fB_0 := C^*_0(\CE,\fA)$.  We also
 denote this as  $\CO^{\rm QMNA}$  when $\CE,\fA$ are understood.
(The restriction to the subalgebra $C^*_0(\CE,\fA)$ is an annoying technicality
required by the desire to use Morita equivalence in the formulation of the Born
rule in section \ref{sec:QMNA-BR}.)

Generalizing the set of states $\CS$ in a satisfactory manner is
more involved. Some motivation is provided by recalling the
standard definition of a \emph{state} in the sense of $C^*$-algebra
theory.

Recall first that an element $a\in \fA$ is \emph{positive} if it can be
expressed in the form $a = b^*b$ for some $b\in \fA$. Denote the
set of positive elements in $\fA$ by $\fA_+$. It is a closed convex cone.
A $\IC$-linear map $\varphi: \fB \to \fA$ between $C^*$ algebras is said to be
\emph{positive} if $\varphi$ takes positive elements of $\fB$ to positive
elements of $\fA$: $\varphi: \fB_+ \to \fA_+$. If $\fB$ and $\fA$ have a
unit a map is \emph{unital} if $\varphi(\textbf{1}_{\fB}) =\textbf{1}_{\fA}$.
Then we have

\bigskip
\noindent
\textbf{Definition}:   A \emph{state}
on a $C^*$-algebra $\fA$ is a positive unital map
$\omega: \fA \to \IC$.
(If $\fA$ is not unital then we take a
positive map of norm one. )

\bigskip
\bigskip

Because the word ``state'' carries several meanings in this paper we will
refer to a ``state'' in the sense of $C^*$-algebra theory as a ``$C^*$-state.''
We will denote the set of $C^*$-states on $\fA$ as $\CS(\fA)$. It is a convex
compact subset of the continuous dual $\fA^\vee$ in the $w^*$ topology.
 Two standard examples of $C^*$-states will play an important role in our story:

\bigskip
\noindent
\textbf{Example 1}: If $\fA = \CB_0(\CH)$ is the $C^*$-algebra of
compact operators on a Hilbert space then the most general $C^*$-state
on that algebra is indeed of the form
\be
\omega(a) = \Tr_{\CH} \rho a
\ee
for some traceclass positive operator $\rho$ of trace one.
\footnote{If we replace ``compact operators'' by ``bounded operators'' this is
no longer true when the Hilbert space $\CH$ is infinite-dimensional. See the remark at the beginning of section 2.13 of
\cite{Landsman}. The states we are describing are known as ``normal states.''
See \cite{KadisonRingrose}, Theorem 7.1.12. }
Thus the
$C^*$-algebra notion of ``state'' coincides with the standard use of the
term ``state'' in quantum mechanics.

\bigskip
\noindent
\textbf{Example 2}: If $\fA$ is commutative, so that
$\fA = C(X)$ is the $C^*$-algebra of
complex-valued functions on a Hausdorff topological space $X$,
then the Riesz-Markov theorem says that the most general
state is   of the form
\be
\omega(f) = \int_{X} f(x) d\mu_x
\ee
for some positive measure $d\mu_x$ on $X$ of total weight one.
Thus, a $C^*$-state on $C(X)$ is the same thing as a classical
probability distribution on $X$. We will denote the $C^*$-algebra
state on $C(X)$ corresponding to the measure $\mu$ by $\omega_\mu$.
\bigskip

A natural generalization of the $C^*$-algebra notion of a state makes use of
the notion of a \emph{completely positive} (CP) map.
\footnote{CP maps are discussed in many textbooks on quantum information
theory. See also \cite{Paulsen}. }
A CP map  $\varphi: \fB \to \fA$ has the property that  for all positive $n\times n$ matrices $ b^*_{ki}b_{kj} \in M_n(\fB)$
(with sum on $k$ understood) and vectors
$a_i \in \fA$ we have
\be\label{eq:UsefulCP}
\sum_{i,j} a_i^* \varphi(b^*_{ki}b_{kj}) a_j \geq 0.
\ee
If a CP map is unital, so $\varphi(\textbf{1}_{\fB}) = \textbf{1}_{\fA}$ we say it is a CPU map.

\bigskip
\noindent
\textbf{Definition}: Let $\fB = C^*(\CE, \fA)$ be the $C^*$-algebra of
adjointable operators on a Hilbert $C^*$-module $\CE$ over a unital $C^*$-algebra $\fA$.
A \emph{QMNA state} for $(\CE, \fA)$ is a CPU map
\be
\varphi: \fB \to \fA.
\ee
We will denote the space of QMNA states by  $\CS(\CE, \fA)$ where it is understood that
$\CE$ is a Morita equivalence bimodule
\footnote{Here by ``Morita equivalence'' we mean ``strong Morita equivalence''
in the sense of $C^*$ algebra theory, rather than algebraic Morita equivalence
defined by the equivalence of the category of modules. Rieffel's theorem shows that
strong Morita equivalence implies algebraic Morita equivalence. For more discussion
see \cite{Landsman}.}
for $\fB_0= C^*_0(\CE,\fA)$ and $\fA$.
We sometimes write $\CS^{\rm QMNA}$ when $\CE$ and $\fA$ are understood.
\bigskip
\bigskip

\textbf{Remarks}

\begin{enumerate}

\item Physical examples make it clear that the definition of a $C^*$-state should be extended to
a map $\fA \to \IC$ which is either $\IC$-linear or $\IC$-anti-linear, although the latter
case is usually not discussed in texts on functional analysis. We will henceforth
take $\omega$ to be linear or anti-linear (it must still be positive and unital).   On the other hand,
in the definition of a QMNA state we must take the map $\varphi$ to be $\IC$-linear
because of complete positivity.
For example, the map of complex conjugation $\CK: \Mat_n(\IC) \rightarrow \Mat_n(\IC)$
is easily seen to be positive but not completely positive (at least if we use tensor
product over the complex numbers).

\item When either $\fA$ or $\fB$ is abelian every positive map $\varphi: \fB \to \fA$ is
completely positive. In particular, if   $\fA=\IC$ so we have a Hilbert space over
a single commutative point then, if $\varphi$ is a normal state,  $\varphi(b)= \Tr_{\CH}(\rho b)$
where $\rho$ is a density matrix, so the definition reduces to the
standard definition of a state when $X$ is a single point - as it must.

\item In the commutative case $\fA = C(X)$, given a Hilbert bundle
$\pi: \CH \to X$ the adjointable operators
$\fB$ are continuous families of operators on the fibers. So
$b\in \fB $ is a continuous section $x \mapsto b_x$ with $b_x \in \CB(\CH_x)$. If we are given a continuous family of density
matrices $\rho_x$  we can define a QMNA state $\varphi$
by declaring $\varphi(b) \in C(X)$ to be the function whose values are given by
\be\label{eq:LocalStates}
\varphi(b)(x)= \Tr_{\CH_x}(\rho_x b_x).
\ee
This is part of the motivation for our definition: It is awkward to
try to generalize separately the section $x \mapsto \rho_x$ and the
traces $\tau_x( \cdots) = \Tr_{\CH_x}(\cdots)$, and it is much
nicer to combine them into a single map $\varphi$ as in
\eqref{eq:LocalStates}.

\item However, it must
be noted that when $\fA = C(X)$, states of the form \eqref{eq:LocalStates} are not the only QMNA states.
Indeed, if $d\mu \in \fM(X)$ and $\rho_{x,x'}$ is any family of
positive traceclass operators (of trace $=1$ ) on $\CH_{x'}$ depending
continuously on  $(x,x') \in X \times X$   then
\be\label{eq:GenCommState}
\varphi(b)(x):= \int_{X}    \Tr_{\CH_{x'}}(\rho_{x,x'} b_{x'}) d\mu(x')
\ee
is a completely positive unital map $\fB \to C(X)$. We expect that there is
a notion of a ``normal QMNA state'' that parallels ``normal $C^*$-algebra state.''
Roughly speaking it should be fiberwise-normal. We expect that
\eqref{eq:GenCommState}  is the most general such state.

\item
%
In the commutative case a pure state can be
associated to a nowhere vanishing section. Then $\langle\Psi, \Psi\rangle$ is
an invertible function on $X$. Therefore, in the general
case we would like to associate a QMNA state to those $\Psi \in \CE$
such that $\langle\Psi,\Psi\rangle^{-1}$ exists.
One way to do this is the following: We know that $\langle\Psi, \Psi\rangle$ is
nonnegative so if it is invertible then we can form
 $\langle \Psi, \Psi\rangle^{-1/2}$. Now we define, for $b\in \fB$:
\be\label{eq:VectorQMNA-state}
\varphi(b):= \langle\Psi, \Psi\rangle^{-1/2} \langle\Psi, b \Psi\rangle \langle \Psi, \Psi\rangle^{-1/2}
= \langle \bar \Psi, b \bar \Psi \rangle
\ee
where $\bar \Psi := \Psi  \langle \Psi, \Psi\rangle^{-1/2}$ has norm one:
$\langle \bar \Psi, \bar\Psi \rangle = \textbf{1}_{\fA}$.
This is completely positive and unital.
\footnote{ There is an associated self-adjoint projector on $\CE$ defined by
$P_{\bar \Psi}(\Phi):= \bar \Psi   \langle \bar \Psi,\Phi \rangle$. It has the
peculiar property that the image of the projector is in general
\underline{not} the line in $\CE$ through $\bar \Psi$.}
We call these \emph{vector QMNA states}. Note that in the
commutative case it reduces to
\be
\varphi(b)(x) = \langle\psi_x, \psi_x\rangle^{-1} \langle\psi_x, b_x \psi_x\rangle = \Tr_{\CH_x}  b_x \frac{ \vert \psi_x \rangle \langle \psi_x \vert}{
\langle \psi_x , \psi_x \rangle }
\ee
and this helps to justify the terminology.

\item A generalization of \eqref{eq:VectorQMNA-state} is
\be
\varphi(b) = \sum_{\alpha} \langle \Psi_\alpha, b \Psi_\alpha \rangle
\ee
where $\{ \Psi_\alpha\} $ is a set of vectors in $\CE$ such that $\varphi(b)$ is a convergent
sum and in particular $\sum_{\alpha} \langle \Psi_\alpha, \Psi_\alpha \rangle_{\fA} = \textbf{1}_\fA$.
As we will soon see, in the finite-dimensional case this is nothing but  the Kraus form of a CPU map.

\item \emph{Purification}: In ordinary quantum mechanics a standard maneuver allows one to replace a density matrix by a pure state on a different Hilbert space. A perfect analog exists in the Hilbert $C^*$-module case, thanks to a theorem of Kasparov
    \cite{Kasparov}.   For each CPU map $\varphi: \fB \to \fA$ Kasparov  constructs a Hilbert module $\CE^{\rm Kasp}$ for $\fA$,
    with $\fB \subset C^*(\CE^{\rm Kasp},\fA)$, and a
    vector $\bar\Psi \in \CE^{\rm Kasp}$ with $\langle \bar\Psi, \bar\Psi \rangle =1 $ so that $\varphi$ becomes a vector
    state \eqref{eq:VectorQMNA-state}. The construction of   $\CE^{\rm Kasp}$ is reviewed in
    appendix \ref{app:Theorems}.

\item \emph{The space of CPU maps}.
In QM the space of states is a compact convex space in the $w^*$-topology.
One can similarly define a $w^*$-topology on   $\CS(\CE, \fA)$ as the weakest topology
such that for all $b\in \fB$ the evaluation maps $\varphi \mapsto \varphi(b) \in \fA$
are continuous. Then
$\varphi_n \to \varphi$ iff for all $b\in \fB$  we have
$\varphi_n(b) \to \varphi(b)$ in $\fA$. It is clear that $\CS(\CE,\fA)$ is closed and
convex in this topology. If $\fA$ is
finite-dimensional then one can imitate the proof of the Banach-Alaoglu
theorem to show that the set of completely positive unital maps
$\fB \to \fA$ is compact. However, an application of the Riesz lemma shows that
the unit ball in a normed linear space is
compact iff that linear space is finite-dimensional
so there is no reason to expect the space of $CPU$ maps $\fB \to \fA$ to
be $w^*$ compact when $\fA$ is infinite-dimensional. Indeed, one can produce a counterexample
to compactness using the Morita equivalence module between $\CA_{1/\theta}$ and $\CA_{\theta}$
described in section \ref{sec:NC-Torus}. (See section \ref{sec:NC-Torus} for the notation.)
One can show that the sequence of CPU maps determined by the
vector states based on $f_n(t) = e^{-r_n t^2} $  with $r_n \to 0$ has the property that
for no subsequence will $\varphi_n(\tilde V)$ converge. It follows that $\CS(\CE,\fA)$ is
convex, and $w^*$-closed, but not $w^*$-compact. This is a significant difference from
standard quantum mechanics.

\item \emph{Extremal States}. Since $\CS(\CE,\fA)$ is a closed convex space we can ask about the
extremal states. In  \cite{BellissardCP} it is shown that the extremal
states are those for which the minimal
Kasparov module $\CE^{\rm Kasp}$ is an ``irreducible'' module for $\fB$.
A module is \emph{minimal} if  $\pi(\fB)\CE$ is dense in $\CE$ and \emph{irreducible} when  any element $T\in C^*(\CE,\fA)$
commuting with $\pi(\fB)$ is a multiple of the identity. It would be very nice to clarify the
relation of between extremal states and vector states in the general case. (In standard QM they are the same.)
In the finite-dimensional case it is easy to show that the extremal states indeed correspond to the vector states:
The minimal module is just $\Hom(\CH_1,\CH_2)$, corresponding to
 a single nonvanishing Kraus operator and hence  are just given by the isometries $E: \CH_1 \to \CH_2$ modulo overall scaling by a phase. This is the $k=1$ space $U(1) \backslash U(d_2)/U(d_2 - d_1)$ in the filtration
 \eqref{eq:CPU-Filt}-\eqref{eq:CPU-Strat}.   (The $U(1)$ is central and can be
 brought to the right.)  It generalizes
the identification of the space of pure states with projective space.

\item Morita equivalence is a reflexive relation so one can in principle exchange the base and fiber in our set up. Indeed given a bimodule $\CE$ we can define a $\fB_0$-valued sesquilinear form $\langle \cdot, \cdot \rangle_{\fB}$ (linear in the first, and anti-linear in the second argument)
such that the crucial identity
\be\label{eq:MoritaIdentity}
\langle \Psi_1, \Psi_2\rangle_{\fB} \Psi_3
= \Psi_1 \langle \Psi_2, \Psi_3 \rangle_{\fA}
\ee
is satisfied. Conversely, given $\CE$ and sesquilinear forms such that \eqref{eq:MoritaIdentity} holds we have a
Morita equivalence between
$\fA$ and $\fB_0$. (See, for example, \cite{Landsman}, Proposition 3.4.4.)

\item An important attribute of a state $\rho$ in QM is the
von Neumann entropy $S(\rho):= - \Tr(\rho \log \rho)$. It is natural to
ask what the generalization is to a QMNA state. A natural guess,
at least in the finite-dimensional case, is that it is either the
channel capacity, or the Holevo chi, discussed in \cite{Holevo,Keyl,NielsenChuang,OhyaPetz,PreskillQI,Preskill:2016htv,Wilde},
of the corresponding quantum channel, but these quantities are not additive under
tensor product \cite{Hastings,CapacityNotAdditive,Preskill:2016htv,Wilde}. A more promising possibility
appears to be squashed entropy \cite{SquashedEntropy,SquashedEntropy2}.
\footnote{I thank A. Kitaev for bringing these papers to my attention.}
However an investigation of whether this proposal is interesting or useful is outside the
scope of this paper.

\end{enumerate}

\section{The Born Rule For QMNA}\label{sec:QMNA-BR}

We now generalize the Born rule to the case of QMNA. It might be worth
pausing and reflecting on why this is not completely straightforward.
First of all, if we mindlessly replace a vector $\psi$ in Hilbert space
by a vector $\Psi \in \CE$ then the value of an expression such as \eqref{eq:QM-text}
would be valued in $\fA$. One would need to assume the existence of
a suitable trace (which might or might not exist) to obtain a nonnegative
real number. Moreover, the expression would not obviously be invariant
under unitary change of basis of the $\psi_\lambda^a$. Finally, there is no
spectral theorem for abstract $C^*$-algebras - a point we will return to below.

In order to overcome the difficulties we have just enumerated
we will try to generalize the Born rule \eqref{eq:Family-BR-1} from
the example of continuous families of quantum systems.  The main
 difficulty we must face is that equation \eqref{eq:Family-BR-1} makes
explicit reference to points $x\in X$. To motivate the pointless generalization we seek,
let us note that the Born rule for continuous families in the commutative case
gives a map
\be
 \CS\times\CO  \times \fM(X) \to \fM(\IR)
\ee
where $\fM(X)$ is the set of positive unit volume measures on $X$:
\be
BR(\rho,T,\mu)(m) = \int_X \Tr_{\CH_x} (\rho_x P_{T_x}(m) ) d\mu_x
\ee
This is  the same information as \eqref{eq:Family-BR-1} because we can
recover the map \eqref{eq:Family-BR-1} by taking Dirac measures localized at points.

Now we recall the second example of $C^*$-states from section
\ref{sec:ObsStates} (the Riesz-Markov theorem). This identifies $\fM(X)$
with the set of $C^*$-algebra states on $C(X)$, so that we can
interpret the Born rule for the commutative case as  a mapping
\be\label{eq:CommBR}
 \CS \times \CO \times \CS(C(X)) \to \fM(\IR)
\ee
where $\CS$ is the space of continuous families of density matrices, while
$\CS(C(X))$ is the set of $C^*$-algebra states on the $C^*$-algebra $C(X)$:
\be
BR( \rho,T,\omega):=  BR( \rho,T,\mu_{\omega}).
\ee

If we view the Born rule in the commutative case as a map \eqref{eq:CommBR}
then the generalization to the noncommutative case is immediate.   The Born rule should be
a mapping
\be
  \CS^{\rm QMNA} \times\CO^{\rm QMNA} \times \CS(\fA) \to \fM(\IR)
\ee
From this viewpoint there is a very natural formula for the Born rule:
\be\label{eq:NaiveBornRule}
\BR(\varphi,T,\omega)(m) = \omega \circ \varphi(P_T(m))
\ee
(Of course, there should be a corresponding version for expectation values
such that $\BR(\varphi,T,\omega)(m)= \EV(\varphi, P_T(m), \omega)$. )

For example, for   the general commutative family $\pi: \CH \to X$ for a commutative
space $X$, the expectation value we would associate to a section $x \mapsto T(x)$
of $\Gamma(\End(\CH))$ in the general QMNA state of the type
\eqref{eq:GenCommState} is
\be\label{eq:GenCommEV}
\EV(T,\varphi,\omega) = \int_X  d\mu_{\omega}(x) \int_X d \mu_{\varphi}(y)  \Tr_{\CH_y} \rho_{x,y} T(y).
\ee

If we want to define the Born rule in full generality then the  formula \eqref{eq:NaiveBornRule} has a technical difficulty:
When $T$ has a finite spectrum equation \eqref{eq:NaiveBornRule}  makes perfectly good
sense because then $P_T(m)$ is just a polynomial in $T$,
and hence an element of the $C^*$ algebra $\fB$. However,
in general \emph{there is no spectral theorem for general
$C^*$-algebras}. Technically the problem is that for a self-adjoint
element $b\in \fB$ we can define $f(b)$ for any continuous function $f$
of a single variable, but to define a projection-valued measure we
need to use functions $f$ such as Heaviside step functions (or more
generally characteristic functions of a measurable subset of $\IR$)
and in general $f(b) \notin \fB$ for such functions.
In order to define the value of $BR(T,\varphi,\omega)$
on Borel subsets $m\subset \IR$ we need some kind of projection-valued
measure. The data $\omega$ solves this problem
rather nicely, at least when it is subject to a ``factorization condition''
described near equation \eqref{eq:FactorCondition} below.
Our construction of the Born rule given the factorization condition on $\omega$
is the following:

Via the GNS construction we have a representation
of $\pi_{\omega}: \fA \to \CB(\CH_{\omega})$ as an algebra of bounded operators on
a Hilbert space. Now, it follows rather trivially given the GNS construction that we have
the diagram:
\be
\xymatrix{
\fA \ar[r]^{\pi_{\omega}} \ar[rd]_{\omega} & \CB(\CH_{\omega})\ar[d]^{ \langle\Omega_{\omega}, ~  \cdot ~ \Omega_{\omega}\rangle } \\
 & \IC \\
}
\ee
where $\Omega_{\omega}$ is the cyclic vector generating the GNS space $\CH_{\omega}$.
(Here and below all diagrams are commutative.)

 But now $\CE$ serves as a Morita equivalence bimodule between $\fA$
and $\fB_0$
and by the Rieffel imprimitivity theorem there is an induced
representation $\pi^{\omega,\CE}: \fB_0 \to \CB(\CH^{\omega,\CE})$. Now,
given an observable $T$ we can consider the image $\pi^{\omega,\CE}(T) \in \CB(\CH^{\omega,\CE})$.
If $T$ is self-adjoint then the image will be self-adjoint and there will be a PVM
$P_{\pi^{\omega,\CE}(T)}$ acting on $\CH^{\omega,\CE}$.

Next, we need to assume that $\pi_\omega\circ \varphi: \fB_0 \to \CB(\CH_\omega)$ factors through the map
$\pi^{\omega,\CE}$ so that we can define $\varphi^{\omega,\CE}$ such that:
%
%
%
%
\be\label{eq:FactorCondition}
\xymatrix{
\fB_0 \ar[r]^{\pi^{\omega,\CE}}\ar[rd]_{\varphi}
& \pi^{\omega,\CE}(\fB_0) \ar[rd]^{\varphi^{\omega,\CE}} & \\
& \fA \ar[r]_{\pi_\omega} & \CB(\CH_{\omega} ) \\
}
\ee
We will call the existence of $\varphi^{\omega,\CE}$ making this
diagram commute the ``factorization condition.''
A sufficient criterion for it to hold is that
\be
\ker \pi^{\omega,\CE} \subset \ker ( \pi_{\omega}\circ \varphi)
\ee
Since there are plenty of states $\omega$ so that $\pi^{\omega,\CE}$ is
faithful this does not appear to be an extremely restrictive condition.

Now the map
\be
 \varphi^{\omega,\CE}: \pi^{\omega,\CE}(\fB_0) \to \pi_{\omega}(\fA)
\ee
is defined on a $*$-invariant norm-closed subspace of $\CB(\CH^{\omega,\CE})$ containing
the unit.  The map is completely positive
and hence by \cite{Arveson1} it has a completely positive extension to a bounded linear map
\be
\widehat{\varphi}^{\omega,\CE}: \CB(\CH^{\omega,\CE}) \to \CB(\CH_{\omega})
\ee
Thus we have the diagram:
\be
\xymatrix{
\fB_0 \ar[r]^{\pi^{\omega,\CE} }\ar[rd]_{\varphi}
& \pi^{\omega,\CE}(\fB_0) \ar@{^{(}->}[r]\ar[rd]_{\varphi^{\omega,\CE}}
&  \CB(\CH^{\omega,\CE} )\ar@{.>}[d]^{\widehat{\varphi}^{\omega,\CE} }
& \\
&
  \fA\ar[r]_{\pi_{\omega}}\ar[rd]_{\omega}
& \CB(\CH_{\omega})\ar[d]^{ (\Omega_{\omega}, ~~ \cdot ~~\Omega_{\omega}) }
&  \\
&
&
\IC
& \\
}
\ee

The existence of the extension    $\widehat{\varphi}^{\omega,\CE}$ ultimately relies on the
Hahn-Banach theorem
 so we have no right to expect it to be unique. However, since it is bounded it is continuous
in the norm topology, and hence in the weak topology and hence the extension to the
weak closure
$\overline{\pi^{\omega,\CE}(\fB_0)}^{\rm wk} $ is unique. This unique extension
 can be applied  to the PVM $P_{\pi^{\omega,\CE}(T)}$. Thus we can   finally set:

\be\label{eq:BR-NC}
\BR(\varphi,T,\omega)(m):= \left\langle \Omega_{\omega}, \widehat{\varphi}^{\omega,\CE}\biggl(P_{\pi^{\omega,\CE}(T)}(m)\biggr) \Omega_{\omega} \right\rangle.
\ee

Equation \eqref{eq:BR-NC} is our central definition. Let us verify that it
really is a probability distribution: The real numbers $BR(\varphi,T,\omega)(m)$
are nonnegative, by positivity of the various maps involved.
If $\{ m_i \}$ is a countable disjoint union then we can write:
\be
\begin{split}
\BR(\varphi,T,\omega)(\amalg_i m_i) = &
\left\langle \Omega_{\omega}, \widehat{\varphi}^{\omega,\CE}\biggl(\lim_{n\to \infty} \sum_{i=1}^n P_{\pi^{\omega,\CE}(T)}(m_i)\biggr) \Omega_{\omega} \right\rangle\\
& =\lim_{n\to \infty} \sum_{i=1}^n \left\langle \Omega_{\omega}, \widehat{\varphi}^{\omega,\CE}\biggl( P_{\pi^{\omega,\CE}(T)}(m_i)\biggr) \Omega_{\omega} \right\rangle \\
& = \lim_{n\to \infty} \sum_{i=1}^n BR(\varphi,T,\omega)(m_i) \\
\end{split}
\ee
because the limit $\lim_{n\to \infty} \sum_{i=1}^n P_{\pi^{\omega,\CE}(T)}(m_i)$
converges in the strong topology, and since $ \widehat{\varphi}^{\omega,\CE}$ is
a bounded operator it is continuous in the norm topology and hence certainly
continuous in the strong topology. The same can be said for the state $\omega$.
That justifies the passage of the limits in the above equations.
Finally, for   $m=\IR$ we have
$P_{\pi^{\omega,\CE}(T)}(\IR)=\textbf{1}$ and since $\varphi$ is unital  so is
$ \widehat{\varphi}^{\omega,\CE}$ and since $\omega$ is a $C^*$-algebra state
we have  $\BR(\varphi,T,\omega)(\IR)= 1$.

\bigskip
\noindent
\textbf{Remark}: Actually, the above
discussion could be carried out without making any reference to GNS representations,
and moreover, it is more naturally formulated in terms of $W^*$ algebras. A $W^*$ algebra
is a $C^*$ algebra that is also the dual of a Banach space. $*$-representations in
bounded operators on a Hilbert space are von Neumann algebras, but there can be more than
one representation of an abstract $W^*$ algebra. Nevertheless, there is an abstract spectral theorem for $W^*$ algebras
\cite{Frank}. Now, there is a functor $\CF$ from the category of $C^*$-algebras and $CP$ maps
to the category of $W^*$ algebras and $CP$ maps given by taking the double dual:
$\CF(\fA) := \fA^{\vee \vee}$. Given the abstract spectral theorem for $W^*$ algebras we can
define
\be\label{eq:MainieroVersion}
\BR(\varphi,T, \omega)(m) := \CF(\omega \circ \varphi)(P_{\iota(T)}(m))
\ee
where $\iota: \fA \to \fA^{\vee\vee}$. I thank Tom Mainiero for this important remark.

\section{The Case When $\fA$ Is Finite Dimensional}\label{sec:FiniteDimensional}

As a first example, suppose that $\fA$ is a finite-dimensional $C^*$-algebra.
It is then isomorphic to a finite direct sum of matrix algebras. Heuristically,
the noncommutative space is a finite union of ``fuzzy points.''
\footnote{There are no characters $\chi:M_n(\IC)\to \IC$, so in this sense
there is no point at all to $M_n(\IC)$. On the other hand, given a noncommutative
algebra $\fA$ one can still consider the spectrum of the center $Z(\fA)$ and interpret
$\fA$ as a sheaf over the spectrum of $Z(\fA)$.   In this sense $M_n(\IC)$ corresponds to a
 noncommutative ``thickening'' of a point. We sometimes refer to algebras $M_n(\IC)$ as ``fuzzy points.''}

We begin with the case where $\fA = M_n(\IC)$ for some
positive integer $n$ and take the $C^*$ module to be   $\CE \cong \Mat_{m\times n}(\IC)$,
for some positive integer $m$ (which, we trust, will not be mistaken for a measurable
subset of $\IR$).
It is obviously a left-module for $\fB = \Mat_m(\IC)$ and a right-module for $\Mat_n(\IC)$ and in
terms of $m\times n$ matrices we have the inner products
\be
\langle \Psi_1, \Psi_2 \rangle_{\fA} :=  \Psi_1^* \Psi_2 \in \fA
\ee
\be
\langle \Psi_1, \Psi_2 \rangle_{\fB} :=  \Psi_1\Psi_2^* \in \fB
\ee
 which satisfy \eqref{eq:MoritaIdentity}.

A famous result of Choi and Kraus (it is just a corollary of the Stinespring theorem, see
appendix \ref{app:Theorems}) states that  the most general
unital completely positive map $\varphi: \fB \to \fA$ is of the form \cite{Choi-1,Kraus}
\be\label{eq:GenQMNA-fd}
\varphi(b) = \sum_{\alpha} E_\alpha^* b E_\alpha
\ee
where $E_\alpha: \IC^n \to \IC^m$ and the restriction that $\varphi$ is unital implies
$\sum_{\alpha} E_\alpha^*  E_\alpha = \textbf{1}$.

Now let $T\in \fB$ be a self-adjoint operator $T = \sum_{\lambda}  \lambda P_{\lambda}$,
where the sum runs over the distinct eigenvalues of $T$ and $P_{\lambda}$ is the orthogonal projection
operator onto the subspace of $\IC^m$ of eigenvalue $\lambda$.

Let us work out the probability density on $\IR$ associated to this data. We must first compute
$\CH_{\omega}$.
The general $C^*$-state $\omega$ is of the form
\be
\omega(a) = \Tr  \rho_{\fA} a
\ee
where $\rho_{\fA}$ is a density matrix on $\IC^n$. Then, if the (not necessarily distinct) nonzero eigenvalues of
$\rho_{\fA}$ are $p_i$, $i=1,\dots, s\leq n$ we have
\be
\CH_{\omega} \cong \oplus_{i=1}^s \IC^n
\ee
but with the Hilbert space inner product
\be
\langle \oplus_i z_1^{(i)}, \oplus_j z_2^{(j)}\rangle = \sum_{i=1}^s p_i \langle z_1^{(i)},z_2^{(i)}\rangle_{\rm standard}
\ee
The Morita equivalence bimodule $\CE$ gives the dual representation:
\be
\CH^{\omega}  \cong \oplus_{i=1}^s \IC^m
\ee
again with the inner product weighted as above. The Morita dual representation of $\fB$ is $\pi^\omega(b)=  b\oplus \cdots \oplus b$.
It is now straightforward to work through the definitions and show that
\be\label{eq:FinDimBornRule}
\begin{split}
BR( \varphi, T, \omega)(m) &   = \sum_{\lambda \in m} \sum_\alpha \Tr_{\IC^n} \rho_{\fA} (E_\alpha^* P_{\lambda} E_\alpha ).\\
\end{split}
\ee

We remark that in  the general case when $\fA \cong \oplus M_{n_i}(\IC)$ we simply have a direct sum of the above
construction. Moreover, many of these considerations can probably be generalized to  Hilbert modules over a special
class of $C^*$ algebras known as $H^*$-algebras \cite{Ambrose} in which $\fA$ can be given
a Hilbert space structure. (For such algebras there is a general spectral theorem
\cite{Saworotnow}.)

\subsection{The Truth About Alice And Bob}

In the discussion above we can identify $\CE$ with a tensor product of finite
dimensional Hilbert spaces $\CE  \cong \CH_B \otimes \CH_A^\vee$ where
$\CH_B \cong \IC^m$ and $\CH_A \cong \IC^n$.  Then, comparison of   equation \eqref{eq:GenQMNA-fd} above with, for example,  equation (3.49)
in Chapter 3 of \cite{PreskillQI} reveals that the QMNA state is nothing but the dual of a quantum channel. Using cyclicity of the trace
we can  write instead:
\be
\begin{split}
\BR(\varphi, T,\omega)(m) &  = \sum_{\lambda \in m}   \Tr_{\CH_B}(  \CC(\rho_{\fA} )  P_{\lambda}  ).\\
\end{split}
\ee
This is the formula for the probability of measuring an operator in $\fB$, that is, an operator on $\CH_B$
after a ``quantum channel''
\be
\CC(\rho_{\fA}) = \sum_\alpha E_\alpha \rho_{\fA} E_\alpha^*
\ee
has been  applied to a state $\rho_\fA$ in $\CH_A$.
 So we can rephrase
the Born rule in the language of quantum information theory
and we  have the following dictionary:
\footnote{We stress that this dictionary only applies to finite-dimensional situations.}

 \vskip1in
\begin{center}
\begin{tabular}{|c|c|}
\hline
QMNA  & Standard QM   \\
\hline
Hilbert module $\CE$
&
Hilbert space of the full system $\CH_{\rm B} \otimes \CH_{\rm A} $ \\
 \hline
``Algebra of functions of & Algebra of operators \\
 (noncommutative) control parameters''  $\fA$
& on $\CH_{\rm A}$  \\
 \hline
Observable $T\in C^*(\CE, \fA)$
&
Self-adjoint operator $T\in \End(\CH_{\rm B})$ \\
 \hline
 Measure  on the ``base'' $\omega$
 &
QM state $\rho_\fA$ on $\CH_{\rm A}$ \\
  \hline
QMNA state $\varphi: \fB \to \fA$  & Dual quantum channel $\CC^*$ \\
  \hline
Born rule $\BR( \varphi, T, \omega)$
&
Measurement by Bob of $T$ \\
& in the state $\CC(\rho_\fA)$ prepared by
Alice\\
&  and sent to Bob via quantum channel $\CC$.   \\
\hline \end{tabular}\end{center}
\vskip1in

\subsection{Relation To Standard QM}\label{subsec:QM-Equiv}

In the previous section we interpreted finite-dimensional QMNA
in terms of standard formulae from quantum information theory
using quantum mechanics of open systems.
Since quantum mechanics of open systems can always be reinterpreted in
terms of quantum mechanics of closed systems we should therefore
be able the interpret the Born rule for QMNA in terms of a standard
Born rule for QM. This is indeed the case.

For the following construction it is convenient to introduce bases
$\{ \vert \psi_n \rangle \}$ for  $\CH_B$ and  $ \{ \vert \chi_a \rangle \}$
for $\CH_A$.
Then  every completely positive map $\varphi$ can be written as
\be\label{eq:EquivDM}
\varphi(b) = \Tr_{\CH_{B}}  \rho_{\varphi} ( b\otimes \textbf{1})
\ee
for a suitable density matrix $\rho_{\varphi} $ in $\End(\CH_{B} \otimes \CH_A)$.
To prove this decompose the Kraus operators as
\be
E_{\alpha} = \sum_{n,a} (E_\alpha)_{na} \vert \psi_n \rangle \langle \chi_a \vert
\ee
where $\varphi(b) = \sum_{\alpha} E_\alpha^* b E_{\alpha}$ and
since it is unital  $\sum_{\alpha} E_{\alpha}^* E_{\alpha} = \textbf{1}_{\CH_A}$.
Then one easily checks that
\be\label{eq:rho-phi-def}
\rho_{\varphi} = \sum_{a,b,n,m}  \left( \sum_{\alpha} (E_{\alpha})_{mb} (E_{\alpha})_{na}^* \right)
\vert \chi_a \rangle \vert \psi_m \rangle  \langle\psi_n \vert \langle \chi_b \vert
\ee
satisfies \eqref{eq:EquivDM}. Thus, the dual of a quantum channel can be interpreted as a
standard QM state on the product of input and output Hilbert spaces.
\footnote{In quantum information theory this is known as the ``Choi-Jamiolkowski theorem,''
or simply channel-state duality. It can be generalized to the case where $\CH_A$ is infinite
dimensional but $\CH_B$ is finite dimensional.}
 Note that due to the unital constraint
the density matrix $\rho_{\varphi}$ satisfies the special condition:
\be\label{eq:SpCn}
\Tr_{\CH_B} \rho_{\varphi} = \textbf{1}_{\CH_A}
\ee

Now, in the QMNA Born rule we compute:
\be
\begin{split}
\omega\circ \varphi(P_T) &   = \Tr_{\CH_A \otimes \CH_B } \rho_{\varphi,\omega} (P_T \otimes \textbf{1}) \\
\end{split}
\ee
where we have defined a density matrix:
\be\label{eq:EquivRho}
 \rho_{\varphi,\omega}  := (1 \otimes \rho_A^{1/2}) \rho_{\varphi} (1 \otimes \rho_A^{1/2})
\ee
Note this is clearly Hermitian and positive. To prove it has trace equal to one it is
necessary to use the condition \eqref{eq:SpCn}.

Thus, the QMNA Born rule can always be expressed as a standard Born rule in
finite dimensions.

\section{Is QMNA Really A Generalization Of Quantum Mechanics?}\label{sec:ReallyGeneralize}

We are now in a position to discuss in what sense QMNA might be a generalization of quantum mechanics. The basic datum in QMNA is a $C ^*$– algebra $\fA$ together with a Hilbert module $\CE $. An ``equivalence" to quantum mechanics would be a pair of one-to-one correspondences:
\be
\CS(\CE,\fA) \times \CS(\fA) \leftrightarrow \CS(\CH)
\ee
\be
\CO(\CE,\fA) \leftrightarrow \CO(\CH)
\ee
Such that if $(\varphi, T,\omega) \leftrightarrow (H, \rho)$ then
\be
\BR(\varphi,T,\omega) = \BR(H, \rho)
\ee

In the finite dimensional case we can indeed put a Hilbert space structure on
 $\CE = \Mat_{m\times n}(\IC) \cong \CH:= \CH_B \otimes \CH_A^\vee $. We can then identify $\CO(\CE,\fA)$ with the operators on
$\CH$ of the form $T_B \otimes 1$. Then given $\varphi$ and
$\omega$ we can produce the ordinary quantum mechanical state given in
\eqref{eq:EquivRho} such that the Born rules are identified. In this case rather than a
\underline{generalization} of quantum mechanics on $\CH$ it is a \underline{specialization}
(with a geometrical point of view): only operators commuting with Alice's Hilbert space are allowed,
  and only density matrices of the special form $\rho_{\varphi,\omega}$ of equation \eqref{eq:EquivRho} are allowed.

On the other hand, if we consider infinite-dimensional algebras then
 QMNA is indeed a generalization of quantum mechanics. There is no natural way to associate a Hilbert space to the data $(\CE,\fA)$. It is true that given the additional data needed to formulate a Born rule, i.e. $(\varphi,T,\omega)$,
 we can in fact construct several Hilbert spaces.
\footnote{See appendix \ref{app:Theorems} for some more details about the constructions mentioned in the next few lines.}
First we can construct the
GNS Hilbert space representation $\CH_{\omega}$ of $\omega$.
Second, we can use the Stinespring theorem to construct a Hilbert space representation associated to the completely positive map
$ \pi_{\omega}\circ \varphi$. Third, we can use $\omega $ to construct a Hilbert space completion of $\CE $ using the sesquilinear
form:
\be
\langle \Psi_1, \Psi_2 \rangle = \omega ( \langle \Psi_1, \Psi_2 \rangle_{\fA} ).
\ee
Finally,  we can use the imprimitivity  theorem to construct the Hilbert space $\CH^{\omega,\CE}$. These Hilbert spaces are all different. For example in the finite dimensional case they have different dimensions. Both Hilbert spaces $\CH_\omega$ and $\CH^{\omega,\CE}$ were of course used in the very definition of the Born rule \eqref{eq:BR-NC}. What is unusual here is that the relevant Hilbert spaces depend on the QMNA state $\varphi$  and/or on the ``measure'' $\omega$ on the base.
Thus ``the Hilbert space depends on the state,'' an assertion that might strike some readers as strange.
Of course, in the ordinary $C^*$ approach to quantum mechanics
the GNS Hilbert space $\CH_{\omega}$ depends on the state $\omega$, so to other readers this will not sound at all unusual.
\footnote{By the ``$C^*$ approach to quantum mechanics" we mean that one begins with a C*algebra $\fA $, identifies  the observables as the self adjoining elements of $\fA$, the physical states with the C*states
of $\fA$, and the
Born rule is then $\BR(\omega,T)(m) = \omega(P_{\pi_\omega(T)} (m))$ where the projection
valued measure acts on the GNS Hilbert space for $\omega$. See
\cite{Blanchard:2016muv,Frohlich1} for readable accounts.}

It is interesting to compare with quantum information theory \cite{Holevo,Keyl,NielsenChuang,OhyaPetz,PreskillQI,Preskill:2016htv,Wilde}. In this case a quantum channel (or,  dually,  a completely positive unital map) can be represented as unitary evolution on a larger Hilbert space. (This is just a corollary of the Stinespring theorem.) In quantum information theory this maneuver is known as ``going to the Church of the larger Hilbert space''  \cite{Quantiki}.  From the viewpoint of quantum information theory the ``Church Hilbert space''
depends only on the channel, but not the state. In our construction, the Church depends on the State - something
that is best avoided.

Finally, we note that some interesting no-go theorems have been proven blocking various
generalizations of quantum mechanics.
It is not our purpose to review the various generalizations of quantum mechanics which have been
proposed. Some notable discussions include \cite{Adler,Mielnik:2000ej,Polchinski:1990py,Townsend:2016xqf,Weinberg:1989cm,Weinberg:1989us}.
One interesting no-go theorem is proven by   Kapustin \cite{Kapustin:2013yda}. He gives an axiomatic
description of quantum systems emphasizing the role of symmetry and how it behaves under composition of
quantum systems and concludes that classical mechanics and standard quantum mechanics are the only possible realizations
of his axioms.
To put an important special case of our discussion in the framework of \cite{Kapustin:2013yda}
we could consider the groupoid of Hilbert bundles over symplectic manifolds where the morphisms are bundle morphisms
preserving the Hermitian structure and covering symplectomorphisms. The QMNA generalization would be the
groupoid of pairs $(\CE, \fA)$ with morphisms given by Banach space isomorphisms $\CE \to \CE' $ compatible
with a $C^*$ isomorphism $\fA \to \fA'$. Already in the case with a commutative base there is no obvious
tensor functor (such as demanded in Kapustin's Axiom 6) which satisfies his Axiom 7. Thus, our setup
does not seem to fit the axiomatic system of \cite{Kapustin:2013yda}, although it is possible that this
can be done by viewing the present paper as  a theory of ``noncommutative superselection sectors.''

\section{K\"ahler Quantization Of Control Parameters:
Recovering The Commutative Case In A Semiclassical Limit}\label{sec:Semiclassical}

One very natural way to obtain ``families of quantum systems over a noncommutative
manifold'' is to consider the case of a  Hilbert bundle over a commutative manifold $X$
and then quantize the algebra of functions on $X$.
When $X$ is compact and the fibers of the Hilbert bundle are finite dimensional we are
just discussing a special case of the finite-dimensional systems examined in section
\ref{sec:FiniteDimensional} but even in this case the added geometrical structure
provides insight. For example, one can take a semiclassical limit, allowing an interesting
comparison of the case of noncommutative and commutative base manifolds.

In principle we would like to take $X$ to be any symplectic manifold. However
there is no known quantization procedure of $C(X)$ in such generality.
Therefore, for simplicity we take $X$ to be a compact K\"ahler manifold with K\"ahler form $\omega$
together with an Hermitian very ample line bundle $L \to X$
such that the curvature of the natural connection on $L$ coincides with the K\"ahler form. This
allows for a quantization of $C(X)$ known as Berezin-Toeplitz quantization.
(See, for example, \cite{Schlichenmaier:2010ui} and references therein for a description of
Berezin-Toeplitz quantization.
\footnote{
Analogous notions exist for a broader class of manifolds - compact almost K\"ahler manifoldss and
${\rm Spin}^c$ manifolds \cite{Kirwin}, but we will restrict attention to compact polarized K\"ahler
manifolds for simplicity.}
)
One advantage of this formalism is that   there is a natural set of ``coherent states'' parametrized by points $x\in X$,
which one may think of as approximations to localizing the system to a particular control parameter.
In addition we assume that the Hilbert bundle  is in fact  an Hermitian holomorphic vector bundle.
In this section we will denote it by $\pi: E \to X$ rather than by the symbol $\CH$ used elsewhere.

The question we will answer  here is just this: \emph{Given a continuous family of observables $t\in \Gamma(\End(E))$
given by $x \mapsto t_x$ and a continuous family $x \mapsto \rho_x$ of density matrices on $E_x$   in the commutative case,
produce naturally associated QMNA data $(\CE,\fA, \widehat{T}\in \fB, \varphi\in \CS(\CE,\fA))$
such that we recover the commutative Born rule in a semiclassical limit.} We will not attempt to make precise
``naturally associated'' so the construction below is a little bit artistic.

\subsection{Quantization of $X$}

To define the family of QMNA systems we recall that in geometric quantization we begin with the Hilbert space
\be
\CH_\kappa : = \ker[ \bar \p :  \Omega^{0,0}(L^{\otimes \kappa})   \rightarrow \Omega^{0,1}(L^{\otimes \kappa}) ].
\ee
Here $\kappa$ is a positive integer playing the role of $1/\hbar$, while
$\Omega^{p,q}(L^{\otimes \kappa})$ is the inner-product space of globally defined $\CC^\infty$  sections
$L^{\otimes \kappa}$ with values in $(p,q)$-forms on $X$. We regard $\CH_\kappa$ as a subspace of a the Hilbert space completion of $\Omega^{0,0}(L^{\otimes \kappa})$. Then we set
\be
\fA_\kappa := \CB(\CH_\kappa).
\ee
When $X$ is compact $\fA_{\kappa}$ is isomorphic to a full matrix algebra $\Mat_{N_{\kappa} }(\IC)$ with
\be
N_{\kappa} = \int_X e^{\kappa c_1(L)} {\rm Td}(T^{(0,1)*}X ).
\ee
The Berezin-Toeplitz quantization of the algebra $C(X)$ is  $f \mapsto Q^{(\kappa)}(f)$ where
\be
Q^{(\kappa)}(f):= \Pi \circ M_f\circ \iota.
\ee
Here $\iota: \CH_\kappa \hookrightarrow \Omega^{0,0}(L^{\otimes \kappa})$ is inclusion,
$M_f$ is the multiplication operator on the Hilbert space $\Omega^{0,0}(L^{\otimes \kappa})$
and $\Pi$ is the orthogonal projection onto the closed subspace $\CH_\kappa$.
We can regard $Q^{(\kappa)}$ as a linear map
\be
Q^{(\kappa)}: C(X) \rightarrow \CB(\CH_\kappa) .
\ee
It can be regarded as a quantization map because while
$ Q^{(\kappa)}(f) Q^{(\kappa)}(g) \not= Q^{(\kappa)}(fg)$,
it is true that (See \cite{Schlichenmaier:2010ui}, Theorem 3.3):
\be
\parallel [ Q^{(\kappa)}(f),  Q^{(\kappa)}(g)]  -\frac{1}{\I \kappa}  Q^{(\kappa)}(\{ f, g\} ) \parallel = \CO(\kappa^{-2})
\ee
for $\kappa \to \infty$. While $Q^{(\kappa)}$ is in general not a morphism of algebras it is
unital: $Q^{(\kappa)}(1) = \textbf{1}$.

Conversely there is a map, known as the \emph{symbol map}
\be
\sigma^{(\kappa)}:\CB(\CH_\kappa) \rightarrow   C(X).
\ee
To define it let $\{ s_a \} $, $a=1,\dots, N_\kappa$  be an orthonormal basis for $\CH_{\kappa}$, and $\widehat T \in \CB(\CH_{\kappa})$
with matrix elements $\widehat{T} s_a = \widehat{T}_{ba} s_b $.
Then $\sigma^{(\kappa)}(\widehat{T})$ is the function on $X$ whose value at $x$ is
\footnote{Summation of repeated indices is assumed throughout, but in some formulae the summation
sign is written explicitly for added clarity.}
\be
\sigma^{(\kappa)}(\widehat{T})(x) :=
\frac{ \sum_{a,b}  s_a(x)^* \widehat{T}_{ba} s_b(x) }{ \sum_{c} s_c^*(x) s_c(x) }.
\ee
Here $s_a(x)\in L^{\otimes \kappa}\vert_x$ is the evaluation of the section $s_a$ at $x$. Both numerator and
denominator are valued in $L^{\otimes \kappa}\vert_x \otimes L^{\otimes \kappa}\vert_x^*$ so the ratio is
canonically a function.    The composition
\be
\IB^{(\kappa)}:= \sigma^{(\kappa)} \circ Q^{(\kappa)}: C(X) \rightarrow C(X)
\ee
is known as the \emph{Berezin transform}. It has the property that there is a semiclassical expansion
for $\kappa \to \infty$ so that
$\IB^{(\kappa)}(f)(x) = f(x) + \kappa^{-1} \Delta f(x) + \CO(\kappa^{-2})$, where $\Delta$ is the Laplacian.

A particularly interesting set of ``measures'' on $\fA_\kappa$ are the
coherent states. Intuitively, these will approach $\delta$-functions at points $x\in X$
in the semiclassical limit $\kappa\to \infty$. The simplest (though not the most conceptual)
way to define them is simply as rank one projection operators  in $\fA_\kappa$:
\be
P_x =  \frac{\sum_{a,b}   h(s_a(x), s_b (x))~  s_a \otimes s_b^\vee  }{  \sum_b h(s_b(x), s_b(x)) }
\ee
where the dual $ s_b^\vee $ is defined using the $L^2$ norm on $\CH_{\kappa}$
(so many people would write $s_a \otimes s_b^\vee = \vert s_a \rangle \langle s_b \vert$)  and again
 $s_a(x) \in L_x^{\otimes \kappa}$ is   evaluation of the section at $x$ and $h$ is the
hermitian metric on $L^{\otimes \kappa}\vert_x$.
We denote the corresponding $C^*$-algebra state by $\omega_{\kappa,x}: \fA_\kappa \rightarrow \IC$.
It is simply:
\be
\omega_{\kappa,x}(A) = \sigma^{(\kappa)}(A)(x)
\ee
Note that $\omega_{\kappa,x}(Q^{(\kappa)}(f)) = \IB^{(\kappa)}(f)(x)$.

\subsection{Quantization Of Sections Of $\End(E)$ For A Bundle Of Hilbert Spaces}

Although it is less-thoroughly discussed in the literature, there is a completely parallel
set of constructions for sections of a holomorphic hermitian vector bundle $E\to X$.
\footnote{Some relevant literature is  \cite{Douglas:2006hz}\cite{Keller:2009vj}\cite{Wang},
although these papers unfortunately do not have precisely the results that we need here.}
Let
$E(\kappa):= E\otimes L^\kappa$. Then $\CH_{\kappa}(E)$ denotes the closed subspace of
the Hilbert space of $L^2$ sections of $E(\kappa)$ consisting of the holomorphic sections.
It is finite-dimensional, if $X$ is compact, with dimension
\be
N_{\kappa}(E) = \int_X \ch(E) e^{\kappa c_1(L)} {\rm Td}(X) = r \kappa^d \vol(X) + \CO(\kappa^{d-1})
\ee
where $\vol(X)$ is the symplectic volume of $X$, $d= \dim_{\IC}(X)$,  and $r$ is the rank of $E$.

We define two operations:
\be
Q^{(\kappa)}: \Gamma(\End(E(\kappa))) \rightarrow \End(\CH_{\kappa}(E)) \ee
\be
\sigma^{(\kappa)}:\End(\CH_{\kappa}(E)) \rightarrow  \Gamma(\End(E(\kappa)))
\ee
where $ \Gamma(\End(E(\kappa)))$ are the $\CC^\infty$ sections of $E(\kappa)$ (suitably completed).
We define  $Q^{(\kappa)}(t)$ exactly as in the function case:  We act on a holomorphic section with the endomorphism $t$ and then project back to
the subspace of holomorphic sections. In order to define the symbol map it is again convenient to introduce some bases.
Thus, let $\{ s_I \}$ denote an ON basis of $\CH_{\kappa}(E)$ with  $I=1, \dots , N_{\kappa}(E)$.     Now,  if $\widehat{T} \in \End(\CH_{\kappa}(E))$ we   define:
\be
\sigma^{(\kappa)}(\widehat{T}) (\psi (x)) := r \frac{\sum_J  h(s_J(x), \psi(x))
(\widehat{T}(s_J))(x) }{\sum_K h(s_K(x), s_K(x))}
\ee
where $\psi(x) \in E(\kappa)_x$ and $h$ is the hermitian metric on $E(\kappa)_x$.

If $\widehat{T}$ has  matrix elements
\be
\widehat{T}(s_I )= \widehat{T}_{JI} s_J
\ee
and
we choose an ON  basis $\{ e_{\alpha}(x)\}$ for the fiber $(E\otimes L^\kappa)\vert_x$  (using the Hermitian metric on $E_x$ and $L_x$)
then we can give a local expansion
$s_{I}(x) = s_{\alpha,I}(x) e_{\alpha}(x) $
and $\sigma^{(\kappa)}(\widehat{T})(x)$ has matrix elements:
\be\label{eq:sig-kappa-mtx-elmnt}
\sigma^{(\kappa)}(\widehat{T})_{\alpha\beta}(x)=r \frac{\sum_{I,J} s_{\alpha I}(x) \widehat{T}_{IJ} (s_{\beta J}(x))^*   }{\sum_{\gamma,K} (s_{\gamma K}(x))^* s_{\gamma K}(x)}
\ee
Note that the symbol map is in general \underline{not} unital.  Using  section 5, theorem 1 of \cite{Wang}
one shows that there is an asymptotic expansion in $1/\kappa$:
\be
\sigma^{(\kappa)}(\textbf{1})(x) \sim \textbf{1} - \frac{1}{\kappa} \left( A_1 - \frac{1}{r} \Tr(A_1) \right)
+ \CO(\kappa^{-2})
\ee
where
\be
A_1(x) = \frac{\I}{2\pi} g^{i \bar i} F_{i \bar i} + \half \CR \textbf{1} \in \End(E(\kappa)\vert_x)
\ee
$F_{i\bar i}$ are the components of the natural connection defined by the Hermitian metric, and $\CR$ is the scalar curvature.

We can define an analog of the Berezin transform by the composition
$\sigma^{(\kappa)}\circ Q^{(\kappa)}$:
\be
\IB^{(\kappa)}: \Gamma(\End(E(\kappa))) \rightarrow \Gamma(\End(E(\kappa)))
\ee
In terms of local bases
\be
\begin{split}
\sigma^{(\kappa)}(Q^{(\kappa)}(t))_{\alpha\beta}(x)& = \frac{ r}{\sum_{\nu,I} (s_{\nu I}(x))^* s_{\nu I}(x)}
\int_{X}
\CB_{\alpha\gamma}(x,y) t_{\gamma\delta}(y) \CB_{\delta\beta}(y,x) \Omega(y) \\
\end{split}
\ee
where the vector bundle analog of the Bergman kernel is
\be\label{eq:MatBergKern}
\CB_{\alpha\beta}(x,y) = \sum_{I} s_{\alpha I}(x) (s_{\beta I}(y))^*
\ee
and $\Omega(y)$ is the symplectic volume form.

\subsection{Choice Of Hilbert Module}

Now we need to choose a Hilbert module $\CE_\kappa$ for $\fA_\kappa$.
Unfortunately, $\fA_\kappa$ does not act naturally on $\CH_{\kappa}(E)$ and therefore we choose:
\be
\CE_{\kappa_f, \kappa_b} :=  \CH_{\kappa_f}(E) \otimes \CH_{\kappa_b}^\vee
\ee
Given our choice of a product of spaces of sections there are two conceptually
different $\hbar$'s so we distinguish them by $\kappa_b$ for quantizing $C(X)$ and
$\kappa_f$ for quantizing $\Gamma(\End(E(\kappa_f))$.
When  $X$ is compact both factors are finite-dimensional so this is just a special case of the
choice made in section \ref{sec:FiniteDimensional}, and $\CE_\kappa$  serves as a Morita equivalence bimodule between
$\fA_{\kappa_b}$ and
\be
\fB_{\kappa_f} =  \CB(  \CH_{\kappa_f}(E)).
\ee

\subsection{Observable, State, Measure, And Born Rule}

Now, if $x\mapsto t_x$ and $x\mapsto \rho_x$ is a ``commutative'' continuous family of
observables and density matrices we would like to produce a ``corresponding''   observable and
state for the QMNA framework. There is an obvious choice for the observable, namely
$\widehat{T} = Q^{(\kappa_f)}(t)$ and we will adopt it. The QMNA state must be a CPU map
\be
\varphi_\rho: \fB_{\kappa_f} \rightarrow \fA_{\kappa_b}
\ee
that, in some sense, approaches $x \mapsto \rho_x$ in the semiclassical limit. There are many choices
we could make and we will - somewhat arbitrarily - just make one and explore it. We will take
\be\label{eq:eff1}
\varphi_{\rho}(\hat T):= Q^{(\kappa_b)}(F_{\hat T})
\ee
where $F_{\hat T}$ is the function whose value at $x$ is:
\be\label{eq:eff1b}
F_{\hat T}(x)=\frac{ \Tr_{E(\kappa_f)_x} \rho(x) \sigma^{(\kappa_f)}(\hat T)(x) }{\Tr_{E(\kappa_f)_x} \rho(x)  \sigma^{(\kappa_f)}(\textbf{1})(x)}.
\ee
(The function in the denominator is required so that our map is unital. Note that for sufficiently large $\kappa_f$ it will
be nonzero.)

Since we wish to compare the QMNA Born rule to the commutative case it is natural to choose the coherent states on $x$ as
the measure on $\fA_{\kappa_b}$.

Putting all this together, if   $\hat T$ is of the form $\hat T = Q^{(\kappa_f)}(t)$ where $t \in \Gamma(\End(E) $ and
we also take a coherent state $\omega_{\kappa_b,x}$ for our measure on the base then the QMNA expectation value is:
\be\label{eq:BR-BT-1}
\EV(\hat T, \varphi, \omega_{\kappa_b,x})= \IB^{(\kappa_b)}(F)(x)
\ee
where
\be\label{eq:BR-BT-2}
F(x) = \frac{ \Tr_{E(\kappa_f)_x} \rho(x) \IB^{(\kappa_f)}(t)(x) }{\Tr_{E(\kappa_f)_x} \rho(x)  \sigma^{(\kappa_f)}(\textbf{1})(x)}
\ee
For the corresponding Born rule we have the somewhat more elaborate formula
\be
\BR(\hat T, \varphi, \omega_{\kappa_b,x})(m)= \IB^{(\kappa_b)}(F_{\rho,t,m})(x)
\ee
with
\be
F_{\rho,t,m}(x) =  \frac{ \Tr_{E(\kappa_f)_x} \rho(x) \sigma^{(\kappa_f)}(P_{\hat T}(m))(x) }{\Tr_{E(\kappa_f)_x} \rho(x)  \sigma^{(\kappa_f)}(\textbf{1})(x)}
\ee
For $X$ compact  $P_{\hat T } (m)$ will be a polynomial in $\widehat{T}=Q^{(\kappa_f)}(t)$.

\subsection{Example 1: A Spin-Half Particle Parametrized By A Sphere}\label{eq:SpinHalfSphere}

Let us take    $E= \IC \IP^1 \times \IC^2$.
The very simplest family of ``commutative''  quantum states we can choose is a spin up state, independent of
control parameter:
\be
\rho(x) = \half ( 1+ \sigma^3)
\ee
  Let us take our commutative family of self-adjoint operators to
be the Bott projector onto the spin in the $\hat x$ direction:
\be\label{eq:BottProjTrue}
 t(x) = \half (1 + \hat x \cdot \vec \sigma) = \frac{1}{1 + \vert z\vert^2}
\begin{pmatrix} \vert z\vert^2 & \bar z \\  z & 1 \end{pmatrix}
\ee
where in the second equation we have introduce stereographic coordinates projecting from the north pole $\hat x^3 = +1$
to the complex plane. We will also use standard polar coordinates $(\theta,\phi)$ with $z=e^{\I \phi} \cot(\theta/2)$.

We quantize the base manifold by taking   $L$ to be the hyperplane bundle equipped with
an Hermitian metric so that, in the usual trivialization of $L^{\otimes \kappa}$
on the northern and southern hemispheres we have
\be
h(\psi_1(x), \psi_2(x)) := (1+\vert z\vert^2)^{-\kappa}\psi_1^*(x) \psi_2(x)
\ee
so  the inner product on $\Gamma(L^{\otimes \kappa})$ is given by
\be
\langle \psi_1, \psi_2 \rangle = \int_{\IC} h(\psi_1(x), \psi_2(x)) \omega(x)
\ee
\be
\omega:= \frac{1}{2\pi} \frac{\I dz \wedge d \bar z }{(1+\vert z\vert^2)^2}
\ee
It is well known that the quantization of the algebra of functions gives the algebra of operators on
the spin $j=\kappa_b/2$ representation of $SU(2)$ and
\be
Q^{(\kappa_b)}(\hat x^i) = \frac{J^i }{ j+1}
\ee
where $J^i$ are Hermitian generators of $\fs \fu(2)$ (and we are using physicist conventions for $SU(2)$
representation theory here).

The eigenvalues of $t(x)$ are always $0$ and $1$ and the Born rule in the commutative case is simply
\be\label{eq:ExampleBR1}
\wp_x(\rho,T)(m) =  \delta_1(m) C   + \delta_0(m)S
\ee
where $\delta_{\lambda}(m)$ is the Dirac measure supported at $\lambda$ and it is convenient to define the functions
\be
C:= \cos^2(\theta/2)\qquad S:=\sin^2(\theta/2).
\ee
In other words: The probability to measure a spin up state in the $\hat x$ direction is
$ \cos^2(\theta/2) $ and the probability to measure it in the opposite direction is $\sin^2(\theta/2) $.

Let us now  repeat the computation in the QMNA setting. We can identify $\CH_{\kappa_f}(E)$ with
$V_{\half}\otimes V_{\frac{\kappa_f}{2}}$ where $V_j$ is the spin $j$ representation of $SU(2)$. We
can then write:
\be\label{eq:SymmetricTee}
\widehat{T} = Q^{\kappa_f}(t) = \half \left( 1 + \frac{2\vec J \cdot \vec S }{j+1} \right)
\ee
where $\vec J$ is in the spin $\kappa_f/2$ representation and $\vec S = \half \vec \sigma$ are the
spin generators in the spin $\half$ representation.  The operator $\widehat{T}$ has
eigenvalues $\frac{\kappa_f+1}{\kappa_f+2}$ on the $j^{tot} = j+\half = \frac{\kappa_f+1}{2}$
subspace and $0$ on the $j^{tot} = j-\half = \frac{\kappa_f-1}{2}$ subspace. It is then
not difficult to check that the QMNA analog of \eqref{eq:ExampleBR1} is
\be\label{eq:QMNA-SU2}
\begin{split}
\BR( \varphi_{\rho}, \widehat{T},\omega_{\kappa_b,z} )(m) &=
\delta_{\frac{ \kappa_f +1}{\kappa_f + 2} }(m) \cdot \Biggl[ \frac{(\kappa_b+1)(\kappa_f+1) + 1}{(\kappa_b+2)(\kappa_f+1)} C
+ \frac{\kappa_b + \kappa_f + 2}{(\kappa_b+2)(\kappa_f+1) } S \Biggr] \\
& + \delta_{0 }(m) \cdot \Biggl[ \frac{(\kappa_b+1)\kappa_f}{(\kappa_b+2)(\kappa_f+1)} S
+ \frac{ \kappa_f }{(\kappa_b+2)(\kappa_f+1) } C \Biggr]. \\
\end{split}
\ee

\subsubsection{Comparison With Standard QM Interpretation}

 It is interesting to contrast with the quantization of $\IC \IP^1 \times \IC \IP^1$
with line bundle $L \boxtimes L^\kappa$ giving the representation
$V_{\half} \otimes V_{ \frac{\kappa}{2} } $  of $SU(2)$. The natural state to compare with is
\be
\rho =  \half (1 + \sigma^3) \otimes P_{z}
\ee
where $P_z$ is the coherent state in the spin $\frac{\kappa}{2}$ representation. The
resulting Born rule for $\widehat{T}$ is
\be\label{eq:CohState-SU2}
\delta_{\frac{ \kappa +1}{\kappa + 2} }(m) \cdot \left( C + \frac{1}{\kappa+1} S \right)
 + \delta_{0 }(m) \cdot   \frac{\kappa}{\kappa+1}  S
\ee
This is not the same as \eqref{eq:QMNA-SU2} even in the special case $\kappa=\kappa_b=\kappa_f$.
Of course, by the results of section \ref{subsec:QM-Equiv} one can find QMNA states and measures $\omega$ to reproduce
\eqref{eq:CohState-SU2} but they will not be very natural.

\subsection{Example 2: The Complex Plane }

Although it is somewhat outside of our technical assumptions we could
take the control parameter space $X$ to be the complex plane $X=\IC$ with
symplectic form:
\be
\omega = \frac{\I dz \wedge d \bar z}{2\pi}
\ee
and Hermitian line bundle just the trivial bundle $L^{\otimes \kappa}$ with Hermitian metric
\be
h_{\kappa}(\psi_1(z,\bar z), \psi_2(z,\bar z)) = e^{-\kappa \vert z \vert^2} \psi_1(z,\bar z)^* \psi_2(z,\bar z).
\ee
An ON basis for $\CH_{\kappa}$, which is now the space of $L^2$ holomorphic functions,  is:
\be
\begin{split}
\psi_s  & := \sqrt{\frac{ \kappa^{s+1}}{s!} } z^s \qquad\qquad s=0,1,2, \dots \\
\end{split}
\ee
This can be identified with the usual ON basis associated with the harmonic oscillators
$[b,b^\dagger]=1$ with
\be
b= \sqrt{\kappa} Q^{(\kappa)}(\bar z) \qquad  b^\dagger = \sqrt{\kappa} Q^{(\kappa)}(z).
\ee
The Berezin-transform of a function is
\be
\IB^{(\kappa)}(f)(w,\bar w) = \kappa \int_{\IC} e^{-\kappa \vert w-z\vert^2} f(z,\bar z)  \frac{dxdy}{\pi}
\ee
and the coherent state projector is:
\be
P_w =
e^{-\kappa \vert w \vert^2}  e^{\sqrt{\kappa} \bar w b^\dagger} \vert 0 \rangle
\langle 0 \vert e^{\sqrt{\kappa} w b}
\ee

Now, we take our bundle to be $E = X \times V$. We will take $V$ to be
 a separable infinite-dimensional Hilbert space and think of it as the Hilbert space
representing a Heisenberg algebra of oscillators $a, a^\dagger$.  Again - the latter is
somewhat outside our technical assumptions, but it is interesting to proceed (just take
a limit on the rank of finite-dimensional subspaces of $V$).

An interesting family of observables to consider is a family of operators
with energy eigenvalues
\be
\langle m \vert H \vert n \rangle = \delta_{m,n} E_n(z,\bar z)
\ee
For example, we could take  $H=\omega(z,\bar z) ( a^\dagger a + \half)$ so the energy eigenvalues are
\be
E_n(z,\bar z) = \omega(z,\bar z) (n+\half).
\ee
A short computation reveals that for any measurable function $f$,
\be\label{eq:BTfH}
\begin{split}
\langle m \vert \sigma^{(\kappa_f)}(Q^{(\kappa_f)}(f(H)))(w,\bar w) \vert n \rangle & = \delta_{m,n} \kappa_f  \int e^{-\kappa_f \vert w-z\vert^2}  f(E_n(z,\bar z) ) \omega(z,\bar z)  .\\
\end{split}
\ee
For a measurement of the energy within a range $[E_1, E_2]$ we would substitute
$f(H) \rightarrow \Theta( E_1\leq H \leq E_2) $  where $\Theta(E_1\leq x \leq E_2)$ is the characteristic function.

As the simplest possible example of a quantum state, let us take the commutative family to be
$\rho = \vert 0 \rangle \langle 0 \vert $, independent of $z$.
Then, for a commutative family we would simply have:
\be\label{eq:HO-CF-1}
\BR(\rho,H,z_0)([E_1,E_2])  = \Theta( E_1 \leq E_0(z,\bar z) \leq E_2)
\ee
Here we are asking: ``What is the probability that the groundstate energy lies between $E_1$ and $E_2$
assuming the state $\rho$ is independent of $z$ and in the groundstate?''
Obviously, the answer is that it is $1$ if $z$ is in the region where $E_1 \leq E_0(z,\bar z) \leq E_2$
and zero otherwise. For example, if $E_0(z,\bar z) = \half \vert z\vert^2$ the relevant region is an
annulus in the $z$-plane.

By contrast, in the QMNA picture, if we consider the QMNA state \eqref{eq:eff1}\eqref{eq:eff1b} and use a coherent
state as the measure on the base, then the result is
\be\label{eq:QMNA-compare}
\begin{split}
\BR^{\rm QMNA}(\rho,H,\omega_{z_0} )([E_1,E_2])
& =  \kappa_r  \int e^{- \kappa_r \vert z_0-w\vert^2}
f(E_0(w,\bar w))     \frac{dx' dy'}{\pi}\\
\end{split}
\ee
where $z=x+\I y$ and $w  = x' + \I y'$ and $\kappa_r := \frac{\kappa_b \kappa_f}{\kappa_b + \kappa_f}$.
Somewhat ironically, when the control parameters are classical the
Born rule is a discontinuous function of control parameters, while,
when the control parameters are quantized, the Born rule is a
smooth function of the parameter (of the coherent state).

\subsubsection{Comparison With Standard QM Interpretation}

In the above discussion let us take the simple case  $H= \frac{\vert z \vert^2}{\kappa_0} (a^\dagger a + 1)$.
Then, if we  view $V$ as the quantization of another copy of $\IC$ with coordinate $z_f$  we can
identify $H$ with
$Q^{(\kappa_0)}(\vert z_b\vert^2 \vert z_f\vert^2) $ (where what we have previously called $z$
is now denoted $z_b$, to avoid confusion).  So we would be starting with a product
of phase spaces $\IC_b \times \IC_f$ with classical Hamiltonian
$h = \vert z_b\vert^2 \vert z_f\vert^2$. The quantization of both factors $\IC_b \times \IC_f$ ``at once'' gives the Hamiltonian
\be
\widehat{H} = Q^{(\kappa_b)}(\vert z_b\vert^2) \frac{1}{\kappa_0} (a^\dagger a + 1) = \frac{1}{\kappa_b \kappa_0}
(b^\dagger b + 1) (a^\dagger a + 1)
\ee

The most natural measurement to compare with the QMNA computation is the following:   There are two oscillators:
The fiber is $a^\dagger, a$ and the base is $b^\dagger, b$. We choose a quantum state
$ \vert z_0 \rangle_{b} \otimes \vert 0 \rangle_{a}$ which is a product of a coherent state for the $b$-oscillator
and the groundstate for the $a$-oscillator. Then the probability to measure the energy in the range $[E_1,E_2]$ is
\be
\wp_{z_0}(\rho,\widehat H) =
Tr_{\CH_{b}} P_{z_0} \Theta( E_1 \leq \frac{1}{\kappa_0\kappa_b} (b^\dagger b + 1)   \leq E_2)
\ee
Expanding the coherent state projector in an eigenbasis of $b^\dagger b$ we can write this as:
\be
\wp_{z_0}( \rho,\widehat H) = e^{-\kappa_b \vert z_0 \vert^2}  \sum_{s=0}^\infty \frac{(\kappa_b \vert z_0 \vert^2)^s}{s!}
\Theta(E_1 \leq \frac{s+1}{\kappa_0 \kappa_b} \leq E_2 )
\ee

\begin{figure}[h]
\includegraphics[scale=0.65] 
{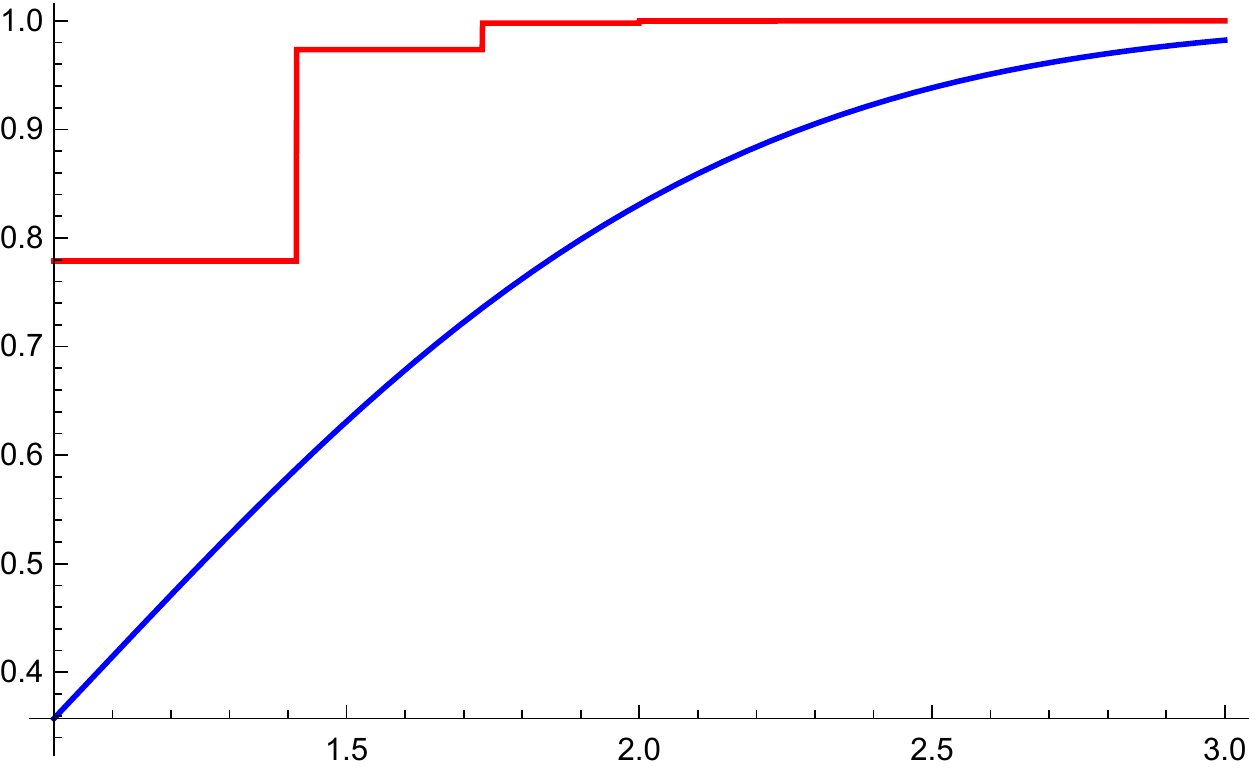}
\caption{Comparing probabilities for the case $E_1=0$, and $r_0=1/2$, as a function of
$e= \kappa\sqrt{E_2}$ for the QMNA (blue) and QM (red) expressions. Both expressions rapidly approach $1$ as
the energy is increased.   }\label{Fig:CompareQM-QMNA}
\end{figure}

\begin{figure}[h]
\includegraphics[scale=0.65] 
{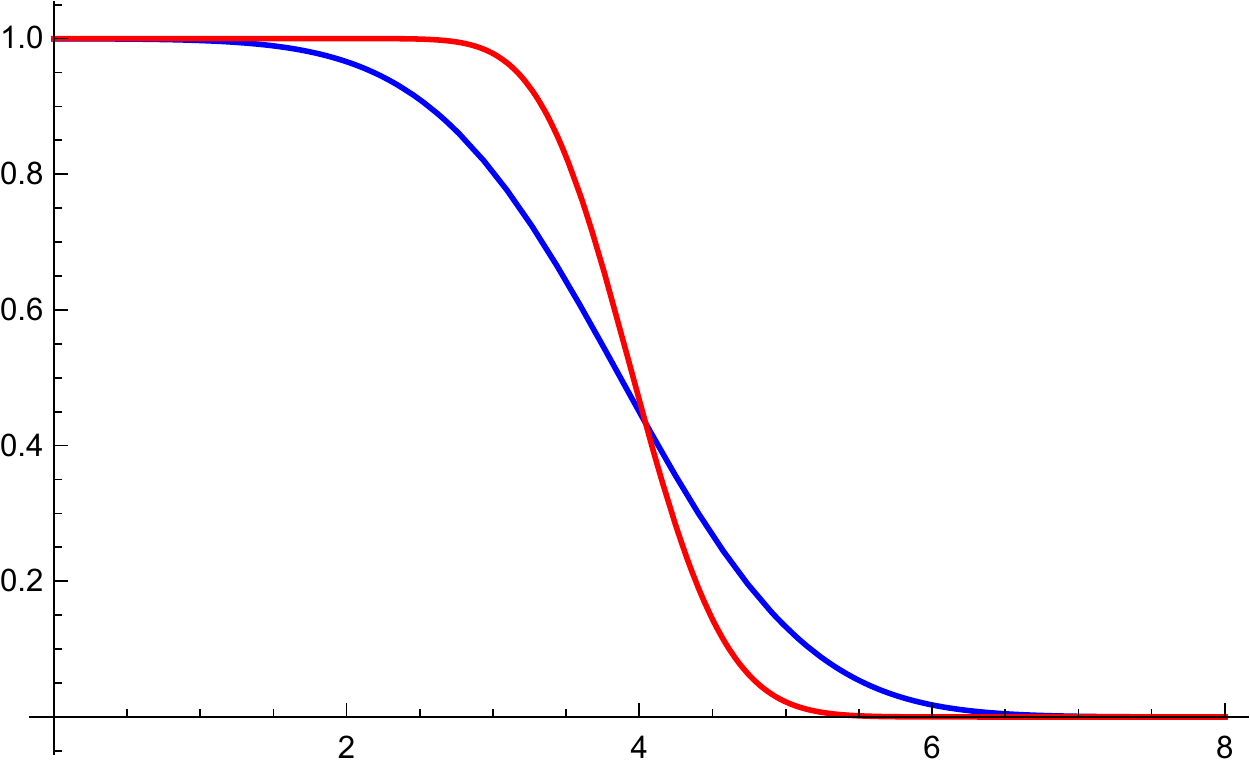}
\caption{Comparing probabilities for the case $E_1=0$,   as a function of $r_0$ for
$e= \kappa\sqrt{E_2}=4$ for the QMNA (blue) and QM (red) expressions. Both expressions rapidly approach $0$ as
the fundamental frequency of the oscillator is increased.   }\label{Fig:CompareQM-QMNA-3}
\end{figure}

Let us compare this with the corresponding QMNA result \eqref{eq:QMNA-compare}. For simplicity
set $\kappa_0 = \kappa_b = \kappa_f = \kappa$. After rescaling variables we have the
QMNA Born rule:
\be\label{eq:QMNA-HO-FAM-1}
e^{-\half r_0^2} \int_{\kappa \sqrt{E_1}}^{\kappa \sqrt{E_2}} I_0(r r_0) e^{-\half r^2} r d r
\ee
where $r_0 = \vert z_0 \vert$. The corresponding QM expression is
\be\label{eq:QMNA-HO-FAM-2}
e^{-r_0^2}  \sum_{s=0}^\infty \frac{r_0^{2s} }{s!}
\Theta(E_1 \leq \frac{s+1}{\kappa^2} \leq E_2 ).
\ee
These are of course not the same. For example, as a function of $E_2$ \eqref{eq:QMNA-HO-FAM-1} is
continuous while \eqref{eq:QMNA-HO-FAM-2} is discontinuous. The two expressions are compared in
Figures \ref{Fig:CompareQM-QMNA}  and \ref{Fig:CompareQM-QMNA-3}.

\bigskip

\subsection{More About The Relation To Standard Quantum Mechanics}\label{subsec:MoreStanQM}

In view of the relation to standard QM explained in section \ref{subsec:QM-Equiv}
  one can compute the effective density matrix \eqref{eq:rho-phi-def} and \eqref{eq:EquivRho} in
  geometrical terms.  We have the finite-dimensional
setup with $\CE = \CH_B \otimes \CH_A^\vee$ with $\CH_B = H^0(E \otimes L^{\kappa_f})$,
with  ON basis $s_I$, $I=1,\dots, N_{\kappa_f}(E)$,  $\CH_A = H^0( L^{\kappa_b})$ with ON basis $s_a$,
$a=1,\dots, N_{\kappa_b}$.  Of the many possible QMNA states we will take
\eqref{eq:eff1}\eqref{eq:eff1b}  and the measure on the base will be a coherent state at $x_0$.
We find that the matrix elements $\rho_{\varphi}$ of \eqref{eq:EquivDM} are
\be
\rho^{\varphi}_{a,J\vert b,I} = \int \frac{ s_a^*(x) (s_{\beta, J}(x))^* \rho_{\beta\alpha}(x) s_{\alpha,I}(x) s_b(x)}{
\rho_{\gamma\delta}(x) s_{\delta,K}(x) (s_{\gamma,K}(x))^* } \Omega(x)
\ee
where we have used the local expansion  of $s_I(x)$ in a local ON basis for
$(E\otimes L^{\kappa_f})\vert_x$. Using the coherent state at $x_0$, the effective density matrix is:
\be\label{eq:EffDM-Kahler}
\rho^{\varphi,\omega}_{a,J\vert b,I}  =
(P_{x_0})_{aa'} \rho^{\varphi}_{a',J\vert b',I} (P_{x_0})_{b'b}
\ee
with
\be
(P_{x_0})_{ab} = \frac{ s_a^*(x_0) s_b(x_0)}{ \sum_c  s_c^*(x_0) s_c(x_0) }
\ee
We now apply this to the special case where
 $\widehat{T} = Q^{(\kappa_f)}(t)$
where $t$ is a continuous
section $t \in \Gamma(\End(E \otimes L^{\kappa_f}))$.   Then we can rewrite
the expectation value in the form \eqref{eq:GenCommEV} using the general
``commutative'' QMNA states we found in \eqref{eq:GenCommState}:
\be\label{eq:EquivQM-Comm1}
\EV(\widehat{T} , \varphi, \omega_{x_0} ) = \int_X  d\mu_{\omega_{x_0} }(x)
\int_X  \Tr_{E_y} (\rho_{x,y} t(y) ) \Omega(y)
\ee
where
\be\label{eq:EquivQM-Comm2}
\left(\rho_{x,y}\right)_{\nu,\mu}
= \frac{ \CB_{\nu\beta}(y,x) \rho_{\beta\alpha}(x) \CB_{\alpha\mu}(x,y)}{\Tr_{E(\kappa_f)_x}(\rho(x)\CB(x,x))}
\ee
\be\label{eq:EquivQM-Comm3}
d\mu_{\omega_{x_0} }(x) =
\frac{ \CB(x_0,x) \CB(x,x_0)}{\CB(x_0,x_0)}  \Omega(x)
\ee
where in the definition of $\rho_{x,y}$ we use the matrix Bergman Kernel of \eqref{eq:MatBergKern}, denoted
 $\CB_{\alpha,\beta}(x,y)$, while in the definition of $d\mu_{\omega}(x)$,
$\CB(x,y)$ is the analog ``scalar Bergman Kernel''  for $H^0(L^{\kappa_b})$.
Thus, our QMNA Born rule for quantized control parameters is
the same as a QMNA Born rule for a commutative family, but with a nonlocally
spread out density matrix as in \eqref{eq:EquivQM-Comm1}-\eqref{eq:EquivQM-Comm3}.

\subsection{Relation To The Born-Oppenheimer Approximation}\label{subsec:ToyModel-BO}

A standard topic in physics, known as the Born-Oppenheimer (BO) approximation,
bears some resemblance to the procedure we have discussed in this section.  The motivating example for the BO approximation is
the derivation of  the ground states of molecules and crystals. The phase space for as system of  nuclei and electrons
is  a product of separate phase spaces for the nuclei and for the electrons. Because of the relative energy scales, the electrons are quantized first, and in the leading approximation the phase space coordinates of the nuclei are treated as classical variables. Thus we obtain a commutative family of quantum systems (of the
elctrons) parameterized by the classical phase space coordinates of the nuclei.
 In the next approximation the phase space coordinates of the nuclei are quantized. The Born-Oppenheimer approximation then is a combination of the adiabatic approximation for the electrons and the semi classical approximation for the nuclei.

Generalizing the standard discussion a little bit we can consider
two K\"ahler manifolds $X_1, X_2$ equipped with Hermitian ample holomorphic
line bundles $L_1, L_2$, respectively. We will think of $(X_1,L_1)$ as the
analog of the phase space for the electrons and $(X_2, L_2)$ as the analog of the phase
space for the nuclei.
Abstractly, there are two $\hbar$'s associated with the powers of $L_1$ and
$L_2$ and the quantum system is obtained by quantizing the total space $X_1 \times X_2$ with line
bundle $L_1^{\kappa_1} \boxtimes L_2^{\kappa_2}$, resulting in a Hilbert space
\be
\CH_{\kappa_1, X_1} \otimes \CH_{\kappa_2, X_2}.
\ee
The quantization of classical  observables, such as a hamiltonian $h(x_1,x_2)$ can be obtained in a two-stage
procedure by first considering $H(\hat x_1, x_2) := Q^{(\kappa_1)}(h(\cdot, x_2))$ to produce a section
$\Gamma( \End(\CH))$ where $\CH$ is the   trivial bundle $\CH = X_2 \times \CH_{\kappa_1, X_1}$,
and then, since $\Gamma( \End(\CH)) = \End(\CH_{\kappa_1, X_1})\otimes C(X_2)$ we can obtain
$\widehat{H} =Q^{(\kappa_2)}\left( Q^{(\kappa_1)}(h)\right)$ by quantizing this section of $\End(\CH)$
with $Q^{(\kappa_2)}$.

Note that in this standard procedure, the bundle of Hilbert spaces over $X_2$ is always trivial.
Moreover, in the standard procedure the bundle is not twisted by a power of a line bundle over $X_2$.
Therefore, the procedure discussed in this section is similar to, but differs from, the standard BO approximation.

\noindent
\textbf{Remarks}:

\begin{enumerate}

\item One way to generalize the BO procedure to obtain a nontrivial hermitian holomorphic bundle
$E\to X_2$ is to consider a
suitable bundle of projective spaces $ \IP(V) \rightarrow Z \rightarrow X_2$ whose Dixmier-Douady class
is derived from the topology of $E \to X_2$. The total space of $Z$ is endowed with a K\"ahler
metric  and a holomorphic line bundle $L \to Z$ whose restriction to the fibers is a suitable power
of the hyperplane bundle on $\IP(V)$. The restriction of the curvature of $L$ to the fibers should be
the appropriate multiple of the K\"ahler metric on $Z$, and should be as well a multiple of the Fubini-Study
metric.

\item As a brief aside we note that the general BO procedure for finding the ground state
can be described as follows. Suppose that $Q^{(\kappa_1)}(H(x_1,x_2)) = H(\hat x_1, x_2)$
has a discrete nondegenerate spectrum for all $x_2$ and choose a basis of eigenstates
$\vert \lambda_i(x_2) \rangle$, $i \geq 0$ depending smoothly on  $x_2$. (This cannot always be done.
For example, if the Berry connection has nontrivial Chern classes then it cannot be done.)
We furthermore assume that the groundstate $\lambda_0(x_2)$ is gapped throughout $X_2$.
Next we write an ansatz for the eigenfunctions of
$\widehat{H}:= Q^{(\kappa_2)}(H(\hat x_1, x_2))$ in the form
\be
\Psi = \int_{X_2} \alpha_i(x_2) \vert x_2 \rangle \otimes \vert \lambda_i(x_2) \rangle \Omega(x_2)
\ee
where $\vert x_2 \rangle$ are coherent states on $X_2$. Now recall that
\be
\frac{\langle x \vert Q^{(\kappa)}(f) - f(x) \vert x \rangle}{\langle x \vert x \rangle } = \IB^{(\kappa)}(f)(x) - f(x) =
\kappa^{-1} \Delta f(x) + \CO(\kappa^{-2})
\ee
where $\Delta$ is the K\"ahler Laplacian. We therefore expect that there is an operator version guaranteeing that
the compression into the groundstate $\vert \lambda_0(x_2) \rangle$ is such that
\be
\langle \lambda_0(x_2) \vert \otimes \langle x_2 \vert \left( Q^{(\kappa_2)}(H(\hat x_1, x_2)) - H(\hat x_1, x_2) \right)
\vert x_2 \rangle \otimes \vert \lambda_0(x_2) \rangle
\ee
is $\CO(\kappa_2^{-1})$. Then we get an approximate groundstate of the full system if the coefficients $\alpha_0(x_2)$
of the coherent states on $X_2$ are chosen to be supported near a zero of this quantity.
In   \cite{Hagedorn} this strategy is applied to the problem of groundstates of nuclei.
It is argued there that one can  produce groundstate wavefunctions exponentially close to the
true wavefunctions in the effective expansion parameter $\kappa_{eff} = (m_e/M_n)^{1/2}$ where $m_e$ is the
mass of the electron and $M_n$ is the mass of the lightest nucleon. It would be very interesting to give a
more general and systematic treatment of the above remarks.

\end{enumerate}

\section{The Noncommutative Torus}\label{sec:NC-Torus}

There is a famous $C^*$ algebra known as the ``irrational rotation algebra''
or the ``noncommutative torus algebra''   which has played an important
role in the mathematical development of noncommutative geometry and has been
widely studied by mathematicians.
It also has been applied in physics in many contexts: See
section \ref{subsec:PhysicalInterpret} below for a brief survey.

\subsection{The Algebra $\fA = \CA_{\theta} $}

 $\CA_{\theta}$ is the $C^*$ algebra generated by
two unitary operators (i.e. $U^* = U^{-1}$ and $V^* = V^{-1}$)
satisfying
\be
VU = e^{2\pi i \theta} UV
\ee
To get a $C^*$-algebra we need a norm-completion
using representations. See Wegge-Olsen \cite{WeggeOlsen}, section 12.3.
The general element $a\in \CA_{\theta}$ can be written
\be\label{eq:GenElt}
a=\sum_{m,n\in \IZ } c_{mn} U^m V^n
\ee
with suitable falloff conditions on the coefficients $c_{mn}$.
There is a unique tracial state invariant under the obvious $U(1)\times U(1)$ automorphism
group, defined by
\be
\tau( \sum_{m,n\in \IZ } c_{mn} U^m V^n ) = c_{00}.
\ee

When $\theta = p/q$ with $q\not= \pm 1$ the algebra, while noncommutative, is Morita equivalent to a
commutative algebra. It is the algebra of
 sections of endomorphisms of a trivial rank $q$ bundle over $T^2$. We will
 be most interested in the irrational case.

\subsection{A Choice Of Hilbert Module}

A fascinating Hilbert module to study is the strong Morita equivalence bimodule
between $\fB = \CA_{1/\theta}$ and $\fA = \CA_\theta$. We start with the
pre-Hilbert module  $\CS(\IR)$, the Schwarz space of functions
of rapid decrease.
\footnote{These are complex-valued functions $f(t)$ on the real line
falling off faster than any power. More precisely ${\rm sup}_{t\in \IR} \vert t^\alpha D^\beta f(t)\vert < \infty$
for all nonnegative integers $\alpha,\beta$.}
This module was first studied by Connes and Rieffel \cite{Connes1,Connes2,Rieffel-1,Rieffel-2,Rieffel-3}.

Let $VU=\lambda UV$, $\lambda=e^{2\pi i\theta}$, and
$\widetilde V\widetilde U=\mu \widetilde U \widetilde V$,
$\mu=e^{2\pi i/\theta}$.
We define left and right actions of $\fB$ and $\fA$, respectively,
 on $f\in\CS(\IR)$ via
 \be
 \begin{split}
  (fV)(t)=e^{2\pi it}f(t)\qquad~ &,\qquad
  (fU)(t)=f(t+\theta)\\
  (\widetilde Vf)(t)=e^{-2\pi it/\theta}f(t)\quad &,\qquad
  (\widetilde Uf)(t)=f(t+1).\\
  \end{split}
\ee
Then for functions
$f,g\in\CS(\IR)$ we can define $\fA$- and $\fB$-valued inner products
\be
\begin{split}
  \langle f,g\rangle^{ }_\fA&=\sum_{m,n} \langle f,g\rangle^{ }_\fA(m,n)
	\cdot U^mV^n\\
  \langle f,g\rangle^{ }_\fA(m,n)&=\theta\int_{-\infty}^\infty
	\overline{f(t+m\theta)}g(t)e^{2\pi i(-nt)}\;dt \\
\end{split}
\ee
and
\footnote{Note! With this definition the inner product is $\IC$-antilinear in
the second argument.}
\be
\begin{split}
  \langle f,g\rangle^{ }_\fB&=\sum_{m,n} \langle f,g\rangle^{ }_\fB(m,n)
	\cdot \widetilde U^m\widetilde V^n\\
  \langle f,g\rangle^{ }_\fB(m,n)&=\int_{-\infty}^\infty
	f(t-m)\overline{g(t)}e^{2\pi i(nt/\theta)}\;dt\ . \\
\end{split}
\ee
One can show (\cite{Rieffel-3}, section 2) that
\be
\langle f,g\rangle^{ }_\fB h=f\langle g,h\rangle^{ }_\fA,
\ee
which is the key statement of Morita equivalence. (see   Proposition 3.4.4 of \cite{Landsman}). Note in this equation
the identity is $\IC$-linear in $f,h$ but $\IC$-antilinear in $g$.

\subsection{Observables}

The study of the spectrum of special elements of $\fB$ is a famous and major
topic in mathematical physics. The easiest case is $\tilde V + \tilde V^*$
and $\tilde U + \tilde U^*$ for which the spectrum is fairly trivially
shown to be $[-2,2]$. More complicated operators tend to have a
spectrum which is a Cantor set when $\theta$ is irrational.
For example, the case of
$T = \tilde U+\tilde U^* + \lambda( \tilde V + \tilde V^*)$
is the famous example of the Harper model, the almost Mathieu equation,
the Hofstadter butterfly etc. Here we will simply consider
 $T = \tilde V + \tilde V^*$, the most trivial observable besides the
identity.

\subsection{An Example Of A Probability Distribution In QMNA For $\fA = \CA_{\theta}$ }

We take a QMNA vector state:
\be
\varphi(b) = \langle f, f\rangle_{\fA}^{-1/2} \langle f, b f\rangle_{\fA} \langle f, f\rangle_{\fA}^{-1/2}
\ee
and we make the specific choice $f(t) = e^{-r t^2}$ with $r>0$. It is a
nontrivial fact that $\langle f, f\rangle_{\fA}$ is invertible \cite{Boca}.
Finally, for the measure   $\omega: \fA \to \IC$ we simply take $\omega = \tau$.  Now the probability
associated with $[E_1,E_2]\subset \IR$ is
\be
{\rm BR}(\varphi,T,\omega)([E_1,E_2])=
\tau( \langle f, \chi_{[E_1,E_2]}(\tilde V + \tilde V^*) f\rangle_{\fA}
\langle f, f\rangle_{\fA}^{-1} )
\ee
where $\chi_{[E_1,E_2]}$ is the characteristic function of the interval.
A small calculation shows the  probability distribution can be expressed as:

\be
\begin{split}
{\rm BR}(\varphi,T,\omega)([E_1,E_2]) & =   \sqrt{\frac{2r}{\pi}} \int e^{-2 rt^2}
\chi_{[E_1,E_2]}(2 \cos \frac{2\pi t}{\theta} ) \\
&
\Biggl[ 1 +  \sum_{k=1}^\infty (-1)^k
\sum'_{\vec m, \vec n \in \IZ^k} e^{\I \phi(\vec m, \vec n)} ( \prod_{s=1}^k c_{m_s,n_s} e^{2\pi \I n_s t - 2r \theta m_s t}) \Biggr]dt \\
\end{split}
\ee
where the prime on the sum indicates that we sum over nonzero values of $(\vec m, \vec n)$ and
\be
c_{m,n}= \exp\left[ - \frac{r\theta^2}{2 } m^2 - \frac{\pi^2}{2r} n^2 + \I \pi \theta m n \right]\qquad\qquad
e^{\I \phi(\vec m, \vec n)} = e^{2\pi \I \theta \sum_{1 \leq s < t \leq k} n_s m_t }
\ee

We have presented this computation merely to show that one can actually do explicit computations
of the QMNA Born rule in infinite-dimensional settings.
It would be nice to compare with an analogous computation in ``standard quantum mechanics''
but we have not attempted to do so. (The most promising avenue is to take $\theta$ to be
rational.)

\subsection{Overview Of Physical Interpretations And Applications Of $\CA_{\theta}$ }\label{subsec:PhysicalInterpret}

The noncommutative torus algebra $\CA_\theta$ discussed in
section \ref{sec:NC-Torus} appears in many places in physics.
Here we review some of those applications. Each one provides a potential application of the ideas of this paper although the kinds of questions we address here are not normally discussed in those applications. So it remains to be seen if there are any useful applications of the ideas of this paper.

\begin{enumerate}

\item
$\CA_\theta$ arises very naturally in the physics of two-dimensional electrons
in a uniform magnetic field with irrational number of flux quanta. It is implicit in the
TKNN paper \cite{TKNN,Thouless2} and used explicitly in many discussions of the quantum Hall effect,
most notably in Bellissard's discussion of the effects of disorder on the
quantization of the Hall conductance \cite{BellissardLH,BellissardNCQHE,Connes2}.
In this context it is referred to as the ``noncommutative Brillouin torus.'' (These
considerations can be generalized to many other homogeneous but aperiodic media
\cite{BellissardLH}.)

In the non-interacting electron approximation of a crystal in a Euclidean space $V$ using
a Schr\"odinger operator
invariant under tranlsation by a lattice $\Gamma \subset V$   one can express the
usual Bloch decomposition of the Hilbert space of states of a single electron
in terms of sections of a Hilbert bundle over the Brillouin torus. The Hilbert
space $L^2(V)$ can be identified with the space of $L^2$ sections of a
Hilbert bundle over the dual torus to $V/\Gamma$ (known as the Brillouin torus   $\widehat{T}= V^\vee/\Gamma^\vee$).
Now, following \cite{Gruber}, we  define $L^2(V)$ as a Hilbert module over $C(\widehat{T})$
by using the inner product of the  ``Wannier transforms''
of two wavefunctions.  Next, we observe that for an abelian group $\Gamma$
\be
C(\widehat{T}) \cong C_r(\Gamma)
\ee
where $C_r(\Gamma)$ is the ``reduced group algebra'' -- the representation of $\Gamma$ by
left-translations in $L^2(\Gamma)$. Now, in ``noncommutative Bloch theory''   we want to replace translations by magnetic translations, and we view the $C^*$
algebra generated by these as the twisted group algebra $\CA=C^*_r(\Gamma, \Theta)$ where
$\Theta$ is a cocycle on $\Gamma$.  In particular if $\Gamma \cong \IZ^2$ is the lattice of translations
of an electron in a two-dimensional crystal and the cocycle $\Theta$ is defined by a uniform magnetic field then
 $C^*_r(\Gamma, \Theta)\cong \CA_\theta$.  Then we let $\CH = L^2(E \vert_{\CD})$ where $\CD$ is a fundamental
domain for the $\Gamma$ action on space  (e.g. a unit cell of the lattice) and $E = V \times \IC^2$ is the trivial
bundle and we take
\be
\CE := \CH \otimes \CA.
\ee
Then we have
\be
\langle \psi_1 , \psi_2 \rangle_{\CE} :=  \sum_{\gamma\in \Gamma} \langle U_\gamma \psi_1, \psi_2 \rangle R_\gamma^{\Theta}
\ee
where $U_\gamma$ is the action of magnetic translations on $\psi_1$ and $R_\gamma^\Theta \in C^*_r(\Gamma, \Theta)$
are twisted right-translations.
 The GNS representation of $\CE$ with respect to a canonical trace $\tau$ on $C^*_r(\Gamma, \Theta)$  is
isomorphic to $L^2(E)$ and hence the Hilbert space of a single electron can be   interpreted as a Hilbert module
over the noncommutative torus  in a way that naturally generalizes the usual Bloch theory.
(This construction   generalizes to the $L^2$ sections of an Hermitian vector bundle over a manifold $E\rightarrow M$ with an action of $\Gamma$ on $M$ that lifts projectively to $E$. We can view $L^2(E)$ as a Hilbert module over $C_r(\Gamma; \Theta)$.)
In \cite{Gruber} this construction is used to discuss when self-adjoint operators in $\fB$ have a Cantor spectrum
(such as happens with the Harper-Hofstadter Hamiltonian). Since the Harper-Hofstadter Hamiltonian can also be realized
in optical lattices  \cite{Aidelsburger,Ketterle}    there are potential applications of our ideas to cold atoms.
It would be helpful to have a good
physical interpretation of $C^*$-algebra states $\omega: \fA \to \IC$ other than $\tau$ in this context.

\item The algebra $\CA_{\theta}$ also shows up frequently in discussing
limits of string theory where string field theory degenerates to a field
theory on a noncommutative spacetime. See \cite{Douglas:2001ba,Konechny:2000dp} for reviews.
Although in this paper we have been viewing the noncommutative base as a parameter space, or moduli space, rather than
as a spacetime, the two concepts can be closely related, as,  for example, in M(atrix) theory \cite{Banks:1996vh}.
In any case, the most precise version of the idea is the  Seiberg-Witten
limit of the open string field theory of a D-brane in the presence of a $B$-field
\cite{Seiberg:1999vs}. It is argued in
\cite{Konechny:2000dp,Martinec:2001hh,Seiberg:1999vs} that in this limit, if the D-brane wraps a torus, then
open strings provide an equivalence bimodule between $P\CA_{\theta}P \cong \CA_{1/\theta}$
for a suitable projector $P$. So we can hope that in the SW limit the D0-D2 strings on
a torus with $B$-field turns into the Morita equivalence bimodule $\CS(\IR)$ discussed above.

In general, given a category of open string boundary conditions labeled by $a,b, \dots $ the algebra
of $aa$ strings $\fA = \CH_{aa}$ and $bb$ strings $\fB = \CH_{bb}$ should be - in some sense -
Morita equivalent with the open string states in $\CH_{ab}$ serving as the Morita equivalence
bimodule. Open string states provide QMNA states:  Given a state $\Psi \in \CH_{ba}$ and its Frobenius dual state $\bar \Psi \in \CH_{ab}$
the map of open string operators $\varphi: b \mapsto \bar \Psi b \Psi$ is a QMNA state. (Indeed a variant of this
construction is used in abstract discussions of the ``Cardy condition'' \cite{Moore:2006dw}.) Thus the open strings
can serve as a kind of quantum channel.  Any open string  state in the space of
$bb$ strings can serve as $\omega$. Thus, the SW limit of D-brane open string field theory
provides a context for the application of our Born rule.
It is not the kind of question which is typically posed about D-branes.

\item As mentioned in the Introduction,  the
 noncommutative torus algebra appears in the discussion of
``noncommutative  $tt^*$-geometry'' whose existence is suggested in  \cite{Cecotti:2013mba}.  Again,
it remains to be seen if the considerations of this paper are useful
in that context.

\end{enumerate}

\section{Time Development And Symmetries}\label{sec:TimeDevelopment}

So far we have only considered ``kinematical'' questions such as the formulation of
states, operators, and the Born rule. Now we briefly discuss dynamical issues.
We leave the relativistic generalization for another time.

We expect time-development to be some rule for taking a QMNA  state $\varphi: \fB \to \fA$
and producing a family $\varphi(t): \fB \to \fA$. That is, it should be a map
from time to the space of completely positive unital maps. Moreover, for time-translationally
invariant situations it should involve some group structure.
One might initially think that a  natural way to produce such a family would be to use a Markov process $\Phi(t)$
and take  $\varphi(t) = \varphi\circ \Phi(t)$,
leading to a version of the Lindblad equation. However, this is unsatisfactory since the
evolution will typically only correspond to a
semigroup, and we do not view that as suitable for a complete theory of time-evolution.
\footnote{Since the universe is generally believed to have had a beginning the requirement of having
a group rather than a semigroup is
of course open to debate. Indeed there are serious proposals that Lindblad equations should be
taken to be fundamental descriptions of time evolution \cite{Ghirardi:1985mt}. But we will not
adopt that point of view.}
Rather the viewpoint adopted here is that  time evolution should be a special case of the
action of a symmetry, namely time-translation symmetry.  Therefore, we focus more   generally
on how one can incorporate symmetries in the present framework.

Symmetries should correspond to   automorphisms
of the Born rule. The set of automorphisms of the Born rule form a   well-defined group: They simply consist of a triplet of 1-1 maps:
\be
S_s: \CS^{QMNA} \rightarrow \CS^{QMNA}, \qquad S_o: \CO^{QMNA} \rightarrow \CO^{QMNA}, \qquad S_b: \CS(\fA)  \rightarrow \CS(\fA)
\ee
such that
\be
\BR\left( S_s(\varphi), S_o(T), S_b(\omega)\right) = \BR(\varphi,T,\omega)
\ee
The maps $S_s$ and $S_b$  will determine $S_o$ so we focus on the pairs $(S_s,S_b)$.
The maps are all invertible, by definition, and since the composition of such
maps preserves the Born rule they define a group of automorphisms of the Born rule. This group is much
too large to be useful. We must impose some conditions on $(S_s,S_b)$.  It is
very natural to impose some continuity conditions: $S_s$ and $S_b$ should be (uniformly) continuous in the $w^*$
topology.
\footnote{We assume uniform continuity because this is what is required in \cite{Kadison}.
It means that for any $b \in \fA$ and any $\epsilon>0$ there is a $\delta>0$ and
$a_1, \dots, a_n \in \fA$ so that if $\vert \omega_1(a_i) - \omega_2(a_i) \vert < \delta$
then $\vert S_b(\omega_1)(b) - S_b(\omega_2)(b) \vert < \epsilon$. One can also define a uniform
structure on $\CS(\CE,\fA)$ using the fundamental system of entourages defined by
$N_{\varepsilon, a_1, \dots, a_n } := \{ (\varphi_1, \varphi_2) \vert  \parallel \varphi_1(a_i) - \varphi_2(a_i)
\parallel < \varepsilon \} $.}
 In addition it is natural to assume that these are affine linear:
\footnote{Of course, this is precisely the assumption called into question in the black hole
information paradox.}
\be\label{eq:Affine-Sb}
S_b( t \omega_1 + (1-t) \omega_2) = t S_b(\omega_1) + (1-t) S_b(\omega_2)
\ee
\be\label{eq:Affine-Ss}
S_s( t \omega_1 + (1-t) \omega_2) = t S_s(\omega_1) + (1-t) S_s(\omega_2).
\ee
We will denote the group of automorphisms of the Born rule satisfying these conditions
by $\Aut(\BR)$.

There are some obvious examples of automorphisms of the Born rule:
We will define an \emph{automorphism } of
a $C^*$ algebra $\fA$ to be a $1-1$   complex linear or complex anti-linear map
 $\alpha: \fA \to \fA$ so that
\footnote{Here we depart from standard usage in the $C^*$-algebra literature, where an automorphism
is generally assumed to be complex linear. However, it is clear that in quantum mechanics the more
relevant concept is that it should be $\IC$-linear or anti-linear.  }
\be
\alpha(a_1 a_2) = \alpha(a_1) \alpha(a_2), \qquad  \alpha(a^*) = (\alpha(a))^* , \qquad \alpha(\textbf{1}) = \textbf{1}.
\ee
(One easily checks that these equations imply
$\parallel \alpha(a) \parallel = \parallel a \parallel$.)    Now $(\beta, \alpha) \in \Aut(\fB) \times \Aut(\fA)$
clearly defines an automorphism of the Born rule via:
\be\label{eq:AutTransforms}
S_s(\varphi) = \alpha \circ \varphi\circ \beta^{-1}, \qquad  S_o(T) = \beta(T), \qquad S_b(\omega) = \omega \circ \alpha^{-1}
\ee

If we assume that $S_b$ is $\IC$-linear, uniformly $w^*$-continuous, and affine linear then Theorem 3.3 (or Theorem 3.4)
of \cite{Kadison} guarantees that there is a $C^*$-automorphism of $\fA$ such that $S_b(\omega) = \omega \circ \alpha^{-1}$
(and more generally if there is a weakly-continuous one-parameter group of maps $t \mapsto S_{b,t}$ then there is a weakly-continuous
one-parameter group of automorphisms $t \mapsto \alpha_t$). It is natural to conjecture that the techniques of \cite{Kadison}
can be generalized to prove that we also have $S_s(\varphi) = \alpha \circ \varphi\circ \beta^{-1}$ for a pair of automorphisms
$(\alpha, \beta)$ (or one-parameter group of automorphisms $(\alpha_t, \beta_t)$ given a weakly continuous one-parameter
group $S_{s,t}$).   This does not follow immediately from \cite{Kadison} because Kadison makes use
of the $w^*$-compactness of the space of states $\CS(\fA)$. In any case, given the truth of our conjecture we would have
\be\label{eq:AutBR-conj}
\Aut(\BR) \cong \Aut(\fB) \times \Aut(\fA).
\ee

Conjecture \eqref{eq:AutBR-conj} is closely related to Wigner's
theorem for the following reason: The most general automorphism of $\CK(\CH)$ or $\CB(\CH)$ is inner, that is
$\alpha(T) = U T U^*$ where $U$ is unitary or anti-unitary according to whether $\alpha$ is $\IC$-linear or
anti-linear.
\footnote{For $\IC$-linear maps this result is quite standard.
The proof for the generalization to linear or anti-linear maps is
a straightforward modification: Note that if $\{ P_i \}$ is a collection of orthogonal rank one self-adjoint projectors
(defined, say,  by an ON basis of $\CH$) then $\{ Q_i:= \alpha(P_i)\}$ is a collection of orthogonal self-adjoint projectors
such that $\sum_i Q_i = \textbf{1}$. We claim the $Q_i$ are also rank one. Suppose not. Then for some $i$
the image of $Q_i$ has an ON basis $\{ w_{i,a}\} $, $a=1, \dots, \dim Q_i \CH>1$, so that $Q_i = \sum_a  Q_{i,a}$.
But then $P_i = \sum_a \alpha^{-1}(Q_{i,a})$. The $\alpha^{-1}(Q_{i,a})$ are nonzero and orthogonal. But this
is impossible if $P_i$ has rank one. Choosing two ON bases $\{ u_i \}$  and $\{ v_i \} $ of $\CH$  inducing
the projectors $P_i$ and $Q_i$, respectively, we can define an isometry $W:\CH \to \CH$ by $W(u_i) = v_i$ and extending by
complex (anti-) linearity. Thus, $\alpha(P_i) = W Q_i W^*$. Now we need to show that when $\alpha$ is evaluated
on an arbitrary operator it is still conjugation by some (anti)-unitary operator $U$.  Observe that $\beta:={\rm Ad}(W^{-1})\circ \alpha$ fixes all the projectors $P_i$. Choose a nonzero eigenvector $\xi$
of some $P_i$. Then it is not difficult to show that the map  $T\xi \rightarrow \beta(T)\xi$ is length-preserving for any
$T\in \CB(\CH)$.
(See \cite{WeggeOlsen}, section 1.10.2.) and since $T\xi$ is dense as $T$ ranges over $\CB(\CH)$ there is a well-defined
$\IC$-linear operator  $\tilde U$ with $\beta(T)\xi = \tilde U T \xi$ for all $T\in \CB(\CH)$. But this implies $\beta(T) = \tilde U
T \tilde U^*$ and since this fixes the set of projectors $P_i$ it is a diagonal matrix of phases in the basis $\{ u_i \}$. Putting this together
with $\beta(T) = W^{-1} \alpha(T) W^{-1,*}$ leads to the desired result.   }
Thus once one has established that the general automorphism of the Born rule is of the form $\rho\rightarrow \alpha(\rho)$ and
$T\rightarrow \alpha(T)$ for some automorphism $\alpha$
of $\CB(\CH)$ the Wigner theorem immediately follows. In this sense our conjecture is a generalization of the Wigner theorem. However it must be noted that for more general $C^*$-algebras there can be interesting outer automorphisms. Quantum systems involving such $C^*$-algebras
could then have exotic non-Wignerian symmetries (and in particular it is not obvious that time-evolution would even be described by a
Hamiltonian). Of course this remark applies with equal validity to the standard $C^*$-algebra approach to quantum mechanics.
Some interesting examples of $C^*$-algebras with nontrivial outer automorphisms are the following:

\begin{enumerate}

\item  Algebras of continuous trace type. The simplest of these are
\be
C_0(X, \CK(\CH)) :=\{ f: X \to \CK(\CH): f ~ {\rm continuous} ~~ f \mapsto \parallel f(t) \parallel \in C_0(X)  \}.
\ee
Pointwise inner automorphisms are families $f(x) \mapsto  u(x) f(x) u(x)^*$, where $x \mapsto u(x)$ is continuous.
If we look at automorphisms $\alpha$ commuting with the obvious action of $C_0(X)$
then such automorphisms are classified, up to pointwise inner automorphism, by a class
$\zeta(\alpha) \in H^1(X, \underline{C(\IT)} ) \cong H^2(X,\IZ)$. In fact, for a general continuous trace $C^*$
algebra with spectrum $X$  (roughly speaking -- this is a bundle of $C^*$-algebras over $X$) we have
\be
0 \rightarrow {\rm Inn}(A) \rightarrow {\rm Aut}_{C_0(X)}(A) \rightarrow H^2(X;\IZ)
\ee
and the last arrow is surjective if $A$ is stable. Moreover
\be
0 \rightarrow \Aut_{C_0(X)}(A) \rightarrow \Aut(A) \rightarrow {\rm Homeo}_{\delta(A)}(X)
\ee
where ${\rm Homeo}_{\delta(A)}(X) $ is the group of homeomorphisms of $X$ that preserve the
Dixmier-Douady class $\delta(A) \in H^3(X;\IZ)$. For proofs of the above facts see
\cite{RaeburnWilliams}.

\item The  Cuntz algebras $\CO_n$. These are the abstract
algebras generated by $n$ partial isometries $S_i$, so (with no sum on $i$):
$S_i^* S_i = \textbf{1} $
but $S_i S_i^*$ is not one - it is just a projector. In addition we have the relation
\footnote{It would be nice to have a physical interpretation of this algebra. One way to
interpret it \cite{Evans} is to consider the tensor space built on an $n$-dimensional Hilbert space $\CH$:
$T(\CH) = \IC \oplus \CH \oplus \CH\otimes \CH \oplus \CH \otimes \CH \otimes \CH \oplus \cdots$. It is
like a fermionic or bosonic Fock space except that one does \underline{not} symmetrize or
antisymmetrize. So we drop the commutation relations $[a_i, a_j] =0$ for $i\not=j$ on the
oscilators. Now choose an ON basis $\{ e_i \}$ for $\CH$ and let $S_i$ be the Hilbert hotel shift operator
of tensoring on the left by $e_i$. One checks that $\sum_{i=1}^n S_i^* S_i = 1- \vert 0 \rangle \langle 0 \rangle$,
and if one mods out the algebra of bounded operators $\CB(T(\CH))$ by the ideal of compact operators the result
is a copy of the Cuntz algebra.}
\be
\sum_{i=1}^n S_i S_i^* = \textbf{1}.
\ee
This has a UHF subalgebra $\CF_n$ which is $M_n \otimes M_n \otimes M_n \otimes \cdots$. An example of
an outer automorphism that restricts to an outer automorphism on $\CF_n$ is
simply $u\cdot S_j = \sum_i u_{ij} S_i$ where $u_{ij} \in U(n)$  \cite{ContiSymanski}.

\item The Calkin algebra $\CB(\CH)/\CK(\CH)$. Rather astonishingly there are $2^{2^{\aleph_0}}$ outer automorphisms of
this $C^*$-algebra \cite{PhillipsWeaver}.

\end{enumerate}

\appendix

\section{The Theorems Of Kasparov, Rieffel, and Stinespring}\label{app:Theorems}

In section \ref{sec:ReallyGeneralize} we mentioned that QMNA data allows one to construct several Hilbert spaces and Hilbert modules. In this appendix we explain a few technical details behind those constructions. The constructions   all have the same structure: Given some data one takes an algebraic
tensor product using the data provided, then one quotients by the annihilator
of a sesquilinear semi-definite form and completes, to produce a representation
of a $C^*$ algebra with desired properties.

\subsection{Prototype: The GNS Construction}

Suppose that $\omega: \fA \to \IC$ is a state. Then we define the sesquilinear
form on $\fA$:
\be
\langle a_1, a_2 \rangle := \omega(a_1^* a_2)
\ee
This is only positive semidefinite. However, the annihilator space
\be
\CN_{\omega} = \{ a\vert \omega(a^* a) =0 \}
\ee
is in fact a linear subspace of $\fA$, thanks to the
Cauchy-Schwarz inequality.
Therefore, the form descends to the quotient $\fA/\CN_{\omega}$ where it is
positive definite. Now we take the Hilbert space completion
to define
\be
\CH_{\omega} := \overline{ \fA/\CN_{\omega} }.
\ee
Note that there is a distinguished cyclic vector $\vert \Omega_{\omega} \rangle = [ \textbf{1} ]$ and
the   representation  of $\fA$ is canonically defined by  $\pi_{\omega}(b) \cdot [a] := [ba]$.
One can show that the   representation is irreducible iff $\omega$ is a pure (extremal) state
\cite{Landsman}. It is a good exercise to work out the construction for finite-dimensional
matrix algebras.

\subsection{Stinespring Theorem}

\textbf{Theorem}: Suppose that $\varphi: \fB \to \fA$ is a completely positive map
with $\fA \subset \CB(\CH)$ a subalgebra of the bounded operators on a Hilbert space.
Then there is a representation $\pi^{\varphi}: \fB \to \CB(\CH^{\varphi})$  and a map
\be
V: \CH \to \CH^{\varphi}
\ee
so that
\be
\varphi(b) = V^* \pi^{\varphi}(b) V
\ee
If $\varphi$ is unital then $V$ is a partial isometry: $V^*V = 1$ and $VV^*$ is a projection operator.

\emph{Proof}:  To prove this we consider the algebraic tensor product $\fB \otimes \CH$ and introduce
a semidefinite sesquilinear form
\be
Q^{\varphi}(b_1 \otimes v_1, b_2 \otimes v_2) := \langle v_1, \varphi(b_1^* b_2) v_2\rangle_{\CH}
\ee
It is not completely obvious that this is semi-definite:
\be
Q^{\varphi}( b_i \otimes v_i,  b_j \otimes v_j) = \sum_{i,j} \langle v_i, \varphi(b_i^* b_j) v_j \rangle_{\CH} \geq 0
\ee
To prove this note that for any collection of $b_i \in \fB$ there is a matrix $a_{ij} e_{ij} \in \Mat_n(\fA)$ so that
\be\label{eq:pos-phi-cons}
\varphi(b_i^* b_j) = a_{ki}^* a_{kj}
\ee
This follows by applying complete positivity to the matrix $b= b_i e_{1i} \in \Mat_n(\fB)$.
Using equation \eqref{eq:pos-phi-cons} we can write:
\be
Q^{\varphi}( b_i \otimes v_i,  b_j \otimes v_j) = \sum_{i,j} \langle a_{ki} v_i, a_{kj} v_j \rangle_{\CH} \geq 0.
\ee
Therefore,  by the Cauchy-Schwarz inequality
\be
\CN^{\varphi}:= {\rm Ann}(Q^{\varphi}):= \{ \xi =  b_i \otimes v_i \vert   Q(\xi,\xi)=0 \}
\ee
 is in fact a linear subspace and we can define $\CH^{\varphi}$ to be the completion of the quotient:
\be
\CH^{\varphi}:=  \overline{ (\fB \otimes \CH)/\CN^{\varphi} }.
\ee
Now we can define
\be
V(v):= [\textbf{1}_{\fB} \otimes v]
\ee
and a small computation shows that  $V^*([b\otimes v])=  \varphi(b) v$.

An important special case is the finite-dimensional situation where $\fB = \End(\CH_2)$ and $\fA = \End(\CH_1)$
where $\CH_1$ and $\CH_2$ are finite-dimensional Hilbert spaces with   $\dim_{\IC}(\CH_1) = d_1$ and $\dim_{\IC}(\CH_2) = d_2$. In this case $\varphi$ can be put into Choi-Kraus form as follows: On the one hand, we know
\be\label{eq:StSp-1}
\CH^{\rm Stinespring} \cong \End(\CH_2)\otimes \CH_1/\CN
\ee
On the other hand,   because this is a representation of the simple algebra $\End(\CH_2)$ we must also have
\be\label{eq:StSp-2}
\CH^{\rm Stinespring} \cong \IC^k \otimes \CH_2
\ee
for some $k$. Once we choose $k$ the map is determined by a choice of isometry:
\be\label{eq:k-isoms}
V: \CH_1 \rightarrow \IC^k \otimes \CH_2
\ee
since $\varphi(b) = V^* (\textbf{1}_{k} \otimes b) V$. Now, if $e_\alpha$ is an ON basis
for $\IC^k$ then we can write
\be
V\psi = \sum_{\alpha=1}^k  e_\alpha \otimes  E_\alpha \psi
\ee
where we define the Kraus operator $E_\alpha$ by $E_\alpha \psi := \langle e_\alpha, V \psi\rangle\in \CH_2 $.
One easily computes that
\be
V^* (e_\alpha \otimes \chi ) = E_\alpha^* \chi
\ee
for $\chi\in \CH_2$ and then it is straightforward to check
$ V^* \left(\textbf{1}_{k} \otimes b \right)  V \psi = \sum_{\alpha} E_\alpha^* b E_\alpha \psi$,
so $\varphi(b) = \sum_{\alpha} E_\alpha^* b E_\alpha $.
Note that we could have used any ON basis $e_\alpha$ to produce the same operator $\varphi$. Moreover,
the smallest possible $\CN$ is $\CN=0$ so it follows from \eqref{eq:StSp-1} that the maximal value for $k$
is  the dimension of $\CH_1 \otimes \CH_2^\vee$, namely  $d_1 d_2$. The smaller values of $k$
correspond to limits where some $E_\alpha$ vanish. Therefore the space of CPU maps  is (we assume $d_1 \leq d_2$  here):
\be\label{eq:CPU-dblquot}
U(d_1 d_2) \backslash U(d_1 d_2^2)/U(d_1 d_2^2 - d_1).
\ee

It follows from the above discussion that in the finite-dimensional case
the space of CPU maps, $CPU(\fB,\fA)$,  is an orbifold with boundaries. On the one hand, it is
a compact convex set, and hence homeomorphic to the ball.
\footnote{To prove this note that the space is embedded in the Euclidean space
$\Hom\left( \End(\CH_2), \End(\CH_1)\right)$. Choose an interior point as an
origin and consider the map $f(x) = x/\parallel x\parallel$ mapping the boundary
to a sphere. Then the space is a cone on that sphere.}
Nevertheless, it has an interesting geometric filtration:
\be\label{eq:CPU-Filt}
X_{-1} = \emptyset \subset X_0 \subset X_k \subset \cdots \subset X_{d_1 d_2 -1} \subset X_{d_1 d_2} = CPU(\fB,\fA)
\ee
with
\be\label{eq:CPU-Strat}
X_k \cong  U(k) \backslash U(kd_2)/U(kd_2 - d_1)
\ee
for  $1\leq k \leq d_1 d_2$, where $X_k$ is  the space of maps where the minimal number of nonzero
Kraus operators is at most $k$. The space $X_{d_1 d_2-1}$ is real codimension one, and is the boundary, homeomorphic
to a sphere.

\subsection{Kasparov Theorem}

A theorem of Kasparov \cite{BellissardCP,Kasparov} gives the general form of a completely positive unital
map $\varphi: \fB \to \fA$ between $C^*$ algebras. One can construct
a Hilbert $C^*$ module $\CE^{\varphi}$ for $\fA$, a representation $\pi: \fB \to C^*(\CE^{\varphi}, \fA)$ and a
vector $\xi_0 \in \CE$ so that $\varphi(b) = \langle \xi_0, \pi(b) \xi_0 \rangle_{\fA}$. The idea of the proof
is to construct the Hilbert module $\CE^\varphi$ as a completion of $ \fB \otimes \fA/\CN$ where we define
an $\fA$-valued inner product on $\fB \otimes \fA$ by saying that
\be
\langle b_1\otimes a_1 , b_2\otimes a_2\rangle_{\fA} := a_1 \varphi(b_1^* b_2) a_2
\ee
and then extending by linearity. Again one proves positive semi-definiteness using
\eqref{eq:pos-phi-cons}, itself a consequence of $\varphi$ being completely positive.
As usual, we define $\CN = \{ \xi \in \fB\otimes \fA \vert (\xi,\xi)=0 \} $. To prove it is
a linear space we need the Cauchy-Schwarz inequality in the form
\footnote{To prove this note that $0 \leq \langle \xi_2 a - \xi_1 , \xi_2 a - \xi_1 \rangle$
for any $a\in \fA$. We can assume $\xi_2\not=0$ so then we can apply this to $a = \langle \xi_2, \xi_1 \rangle/ \parallel \xi_2\parallel^2$.}
\be
\langle \xi_1, \xi_2 \rangle \langle \xi_2 , \xi_1 \rangle \leq \parallel \xi_2 \parallel^2 \langle \xi_1, \xi_1 \rangle.
\ee
Clearly $\CN$ is a right $\fA$-module so   $\fB \otimes \fA/\CN$ is a right $\fA$-module, and the $\fA$-valued inner product is
\be
\langle [\sum_i b_i \otimes a_i] , \sum_j [ b_j' \otimes a_j' ]\rangle_{\fA} := \sum_{i,j} a_i^* \varphi(b_i^* b_j') a_j'
\ee
The left $\fB$-action is $\pi(b)\cdot [ b'\otimes a'] := [bb'\otimes a']$ and $\xi_0 = [\textbf{1}_{\fB} \otimes \textbf{1}_{\fA} ]$.
Now it is a straightforward computation to verify  $\varphi(b) = \langle \xi_0, \pi(b) \xi_0 \rangle_{\fA}$.

\subsection{Rieffel Induction And Imprimitivity Theorem}

When we have a Morita equivalence bimodule $\CE$ between $\fA$ and $\fB$
then we have an equivalence of representations: Given a state $\omega$ on $\fA$ defining
a representation $\CH_{\omega}$ of $\fA$
we get another representation $\CH^{\omega,\CE}$ of $\fB$. We can construct it
by completing
\be
\CH^{\omega,\CE} :=\overline{ \CE \otimes \CH_{\omega}/\CN_{\omega,\CE} }
\ee
where  $\CN_{\omega,\CE} ={\rm Ann}(Q^{\omega,\CE})$ and
\be
\begin{split}
Q^{\omega,\CE}\left( \Psi_1 \otimes [a_1], \Psi_2 \otimes [a_2] \right)
& = \omega( a_1^* \langle \Psi_1, \Psi_2 \rangle_{\fA} a_2 ). \\
\end{split}
\ee

This is generalized to arbitrary representations $\CH$ for $\fA$ by forming the representation
$\CH^{\CH,\CE}$ for $\fB$ using the algebraic tensor product $\CE \otimes \CH$ and defining
the sesquilinear form:
\be
Q^{\CH,\CE}( \Psi_1 \otimes v_1, \Psi_2 \otimes v_2):= \langle v_1, \pi( \langle \Psi_1, \Psi_2\rangle_{\fA} ) v_2\rangle_{\CH}
\ee
and as usual taking the quotient by the annihilator $\CN^{\CH,\CE}:= {\rm Ann}(Q^{\CH,\CE})$ and
completing.

The imprimitivity theorem says that given a Morita equivalence bimodule $\CE$ between $\fA$ and $\fB$
the association $\CH \to \CH^{\CH,\CE}$ of representations is an equivalence of categories of $C^*$-algebra
representations.

If $\CE$ is a right-Hilbert module over $\fA$ and we are given a state $\omega: \fA \to \IC$ then
we can turn a quotient of $\CE$ into a Hilbert space by defining
\be
\langle e_1 , e_2 \rangle := \omega(\langle e_1, e_2 \rangle_{\fA} )
\ee
and then completing the quotient $\CE/\CN$ as usual. This gives a Hilbert space
representation of $\fB = C^*(\CE,\fA)$  which we can identify with  $\CH^{\omega,\CE}$.

\subsection{Relations Between The Constructions}

There are a number of interesting relations between the above constructions.
In QMNA we have three pieces of data: A  Morita equivalence bimodule  $\fB \leftrightarrow \CE^{\rm Morita} \leftrightarrow \fA$,
a   CPU map $\varphi: \fB \to \fA$, and  a $C^*$-algebra state $\omega: \fA \to \IC$.
We can now make several constructions and ask how they are related:

\begin{enumerate}

\item By Kasparov, a Hilbert $C^*$ module $\CE^{\varphi}_{\rm Kasparov} \leftrightarrow \fA$
with a map $\pi: \fB \to  C^*( \CE^{\varphi}_{\rm Kasparov}, \fA)$.

\item By GNS, a representation $\CH_{\omega\circ\varphi}^{GNS} $ of $\fB$.

\item By Stinespring, a representation $\CH^{\pi_{\omega}\circ \varphi}_{\rm Stinespring} $ of $\fB$.

\item By Rieffel, a representation $\CH_{\omega}$ of $\fA$ and hence a representation $\CH^{\omega, \CE^{\rm Morita}}_{\rm Rieffel}$ of $\fB$.

\end{enumerate}

So, with this data we have \underline{three} Hilbert space representations of $\fB$
and \underline{two}  Hilbert $C^*$-modules for $\fA$.
These Hilbert spaces and modules will all be related to each other. It is useful to get a picture
of the relations by working out the finite-dimensional case.
 Thus, suppose $\fB = \Mat_{m}(\IC)$ and $\fA = \Mat_{n}(\IC)$. The Morita equivalence bimodule in this
case would be $\CE^{\rm Morita} = \Mat_{m\times n}(\IC)$. Now, suppose we have a CPU map in Kraus form
\be
\varphi(b)= \sum_{\alpha=1}^k E_\alpha^*  b E_\alpha
\ee
with $E_\alpha \in \Mat_{m\times n}(\IC)$, $\alpha = 1, \dots, k$ and $\sum_{\alpha} E_\alpha^* E_\alpha=1$.

We begin by computing the Kasparov module. This is  $\fB \otimes \fA / \CN$ and to figure out $\CN$ we look for elements $b_i \otimes a_i$
so that
\be
\begin{split}
( b_i \otimes a_i, b_j \otimes a_j) & = \sum_{\alpha}  ( \sum_i b_i E_\alpha a_i)^* (\sum_j b_j E_\alpha a_j)=0 \\
\end{split}
\ee
So $b_i \otimes a_i  \in \CN \subset \Mat_{nm\times nm}(\IC)$ iff, for all $\alpha$, the sum
$\sum_i b_i E_\alpha a_i = 0 $ in $\Mat_{m \times n}(\IC)$. We therefore have an exact sequence of
vector spaces
\be\label{eq:ExctSeq}
0 \rightarrow \CN \rightarrow \fB \otimes \fA \rightarrow \oplus_{\alpha=1}^k \Mat_{m\times n}(\IC)
\ee
where the second arrow is defined by $b\otimes a \mapsto \oplus_{\alpha} b E_\alpha a$ so that
\be
\CE^\varphi_{\rm Kasparov} \cong \IC^k \otimes \Mat_{m\times n}(\IC)
\ee
provided we have surjectivity of the second arrow in \eqref{eq:ExctSeq}.
Surjectivity will hold for sufficiently small $k$. In particular, for $k=1$
we get $\CE^\varphi_{\rm Kasparov}= \CE^{\rm Morita}$. On the other hand, by  taking the
set of Kraus operators to be the matrix units in $\Mat_{m\times n}(\IC)$ we can get $\CN=0$.
So, in general the Kasparov $C^*$-module is much bigger than the original Morita $C^*$-module
$\CE^{\rm Morita}$. This is a manifestation of how the ``Church depends on the state''  in QMNA.

We can also compare the three possible Hilbert spaces in this example. Now in addition to
$\CE^{\rm Morita}$ and $\varphi$ we are given   a state $\omega: \Mat_n(\IC) \to \IC$.
WLOG we can assume that there are orthogonal vectors $v_s \in \IC^n$ and $0<p_s\leq 1$ for $1\leq s \leq S\leq n$ with
$\sum p_s =1 $ and   $\omega(a) = \sum_s^S  p_s   v_s^* a v_s$.
The Rieffel-induced representation of $\fB$ will just be
\be
\CH^{\rm \CE^{\rm Morita},\omega}_{\rm Rieffel} \cong \oplus_{s=1}^S \IC^m.
\ee
On the other hand, to determine the GNS representation
 $\CH^{\rm GNS}_{\omega\circ \varphi}$  of $\fB$ associated with $\omega\circ \varphi$ we need just compute
\be
\begin{split}
\omega\circ \varphi(b)& = \sum_{\alpha, s}  p_s (E_\alpha v_s)^* b (E_\alpha v_s) = \sum_{\alpha,s} w_{\alpha,s}^\dagger b w_{\alpha,s}\\
\end{split}
\ee
where $w_{\alpha,s} = \sqrt{p_s} E_\alpha v_s $, for $ 1\leq \alpha \leq k, 1\leq s \leq S$,  is a set of vectors with
 $\sum_{\alpha,s} w_{\alpha,s}^\dagger w_{\alpha,s} = 1$.
We know that there is a set of ON vectors $u_a \in \IC^m$ with
\be
\begin{split}
\sum_{\alpha,s} w_{\alpha,s} w_{\alpha,s}^\dagger &  = \sum_{a=1}^D q_a u_a u_a^\dagger\\
\end{split}
\ee
where  $0 < q_a \leq 1$ and   $D \leq {\rm Min}[k\cdot S, m]$.
Applying the GNS construction we see that
\be
\CH^{\rm GNS}_{\omega\circ \varphi} \cong \oplus_{a=1}^D \IC^m.
\ee
Finally, to compute   $\CH_{\rm Stinespring}^{\pi_\omega\circ \varphi}$ we consider
\be
\begin{split}
Q^{\pi_{\omega}\circ \varphi}(b_x \otimes v_x, b_x \otimes v_x ) &  = \sum_{s=1}^S \sum_{\alpha=1}^k p_s
\langle b_x E_\alpha v_x^{(s)} , b_y E_\alpha v_y^{(s)} \rangle \\
\end{split}
\ee
for $v_x = \oplus_s v_x^{(s)}$.
So now we define a linear map
\be
\fB \otimes \CH_{\omega} \rightarrow \oplus_{s=1}^S \oplus_{\alpha=1}^k \IC^m
\ee
by
\be
b_x \otimes v_x \mapsto \oplus_{s=1}^S \oplus_{\alpha=1}^k b_x E_\alpha v_x^{(s)}
\ee
so that the kernel is exactly $\CN^{\pi_{\omega\circ \varphi}}$. Again, for sufficiently small $k$ the map is onto and
\be
\CH_{\rm Stinespring}^{\pi_\omega\circ \varphi} \cong \oplus_{a=1}^{kS} \IC^m.
\ee

\end{document}